\documentclass[fleqn,preprint,5p]{elsarticle}
\usepackage{times} 
\usepackage{hyperref}
\usepackage{color}
\usepackage{amsmath, amssymb, amsfonts}
\usepackage{graphicx}
\usepackage{wasysym} 
\usepackage{mathrsfs} 
\usepackage{graphicx}  
\usepackage{amsbsy} 
\usepackage{amscd} 
\usepackage{bbm}  % bbm – "Blackboard-style" cm fonts. Blackboard variants of Computer Modern fonts. 
\usepackage{accents} 
\usepackage{multirow} 
 \usepackage{breakurl} % supported command: \burl{url}
\usepackage{epsfig}
\usepackage{array}
\usepackage{cancel}
\usepackage{multicol}
 \usepackage{lipsum}
 \usepackage{cuted}

\journal{Physics Letters B}

\biboptions{numbers,sort&compress}

\def\beq{\begin{equation}}
\def\eeq{\end{equation}}
\def\beqa{\begin{eqnarray}}
\def\eeqa{\end{eqnarray}}
\def\ben{\begin{enumerate}}
\def\een{\end{enumerate}}
\def\bit{\begin{itemize}}
\def\eit{\end{itemize}}

\def\mathswitch#1{\relax\ifmmode#1\else$#1$\fi}
\def\mathswitchr#1{\relax\ifmmode{\mathrm{#1}}\else$\mathrm{#1}$\fi}
\newcommand{\PW}{\mathswitchr W}
\newcommand{\PZ}{\mathswitchr Z}
\newcommand{\PH}{\mathswitchr H}

\newcommand{\Pb}{\mathswitchr b}
\newcommand{\Pt}{\mathswitchr t}

\newcommand{\MW}{\mathswitch {M_\PW}}
\newcommand{\MZ}{\mathswitch {M_\PZ}}
\newcommand{\GZ}{\mathswitch {\Gamma_\PZ}}
\newcommand{\GW}{\mathswitch {\Gamma_\PW}}
\newcommand{\MH}{\mathswitch {M_\PH}}

\newcommand{\mb}{\mathswitch {m_\Pb}}
\newcommand{\mt}{\mathswitch {m_\Pt}}
\newcommand{\mf}{m_f}

\newcommand{\mz}{\mathswitch {\overline{M}_\PZ}}

\newcommand{\gz}{\mathswitch {\overline{\Gamma}_\PZ}}
\newcommand{\as}{\alpha_{\mathrm s}}
\newcommand{\at}{\alpha_\Pt}
\newcommand{\seff}[1]{\sin^2\theta_{\rm eff}^{\rm #1}}

\newcommand{\gev}{\,\, \mathrm{GeV}}
\newcommand{\mev}{\,\, \mathrm{MeV}}

\newcommand{\OO}{{\mathcal O}}

\newcommand{\mycaption}[1]{\caption{\sl #1}}
 
\begin{document}
%\twocolumn
\begin{frontmatter}
 
%\thispagestyle{empty}
%\allowdisplaybreaks

%\begin{flushright}
%DESY 16-119
%\\
%KW 16-002 
%\\
%July 2016      
%\end{flushright}

\title{ Complete electroweak two-loop corrections to $Z$ boson production and decay}
\author[label1]{Ievgen Dubovyk}
\author[label2]{Ayres~Freitas\corref{ca1}}
\cortext[ca1]{Corresponding author} 
\ead{afreitas@pitt.edu}
\author[label3]{Janusz Gluza}
\author[label3]{Tord Riemann}
\author[label4]{Johann Usovitsch}

\address[label1]{
II. Institut f{\"u}r Theoretische Physik, Universit{\"a}t Hamburg,
%Luruper Chaussee 149, 
22761 Hamburg,  Germany} 
\address[label2]{Pittsburgh Particle physics, Astrophysics \& Cosmology Center
(PITT PACC),\\ Department of Physics \& Astronomy, University of Pittsburgh, Pittsburgh, PA 15260, USA}
 \address[label3]{Institute of Physics, University of Silesia, 40-007 Katowice, Poland}
\address[label4]{Trinity College Dublin (TCD) -- School of Mathematics,
Dublin 2, Ireland} 

\begin{abstract}
This article presents results for the last unknown two-loop contributions to
the $Z$-boson partial widths and $Z$-peak cross-section. These are the so-called
bosonic electroweak two-loop corrections, where
``bosonic'' refers to diagrams without closed fermion loops. 
Together with the corresponding results for the $Z$-pole asymmetries $A_l, A_b$, which have
been presented earlier, this completes the theoretical description of 
$Z$-boson
precision observables at full two-loop precision within the Standard Model.
The calculation has been achieved   through a combination of different methods:
(a) numerical integration of Mellin-Barnes representations with contour rotations
and contour shifts to improve convergence; (b) sector decomposition with
numerical integration over Feynman parameters; (c) dispersion relations for
sub-loop insertions. Numerical results are presented in the form of simple
parameterization formulae for the total width, $\Gamma_\PZ$, partial decay widths  
$\Gamma_{e,\mu},\Gamma_{\tau},\Gamma_{\nu},\Gamma_{u},\Gamma_{c},\Gamma_{d,s},\Gamma_{b}$, branching ratios 
$R_l,R_c,R_b$ and the hadronic peak cross-section, $\sigma_{\rm had}^0$. 
Theoretical intrinsic uncertainties from missing higher orders are also discussed. 
\end{abstract}

%\begin{keyword}
%radiative corrections ...
%\end{keyword}

\end{frontmatter}

\section{Introduction}
The number of Z bosons collected at LEP in the 1990's, $1.7 \times 10^7$, together with SLD data made it possible to 
determine electroweak pseudo-observables (EWPOs) with high precision: the $Z$-boson mass $M_\PZ$, its decay width  $\Gamma_\PZ$, branching 
ratios $R$, forward-backward and left-right asymmetries (\rm{or equivalently}  $A_f$ or $\seff{f})$    
 \cite{ALEPH:2005ablast0}. At that time, theoretical calculations, which included complete one-loop Standard Model
corrections, selected higher order QCD corrections, and partial electroweak two-loop results with intricate QED 
resummations, were accurate enough to go hand-in-hand with experimental demands 
\cite{Bardin:1997xq,Bardin:1995XX}. 
However, up to $5 \times 10^{12}$ $Z$-boson decays are planned to be 
observed at projected future $e^+e^-$ machines (ILC, FCC-ee, CEPC) running at the $Z$-boson resonance \cite{Baer:2013cma,Gomez-Ceballos:2013zzn,dEnterria:2016sca,CEPC-SPPCStudyGroup:2015csa}.
These statistics are several orders of magnitude larger than at LEP 
and would  lead
to very accurate experimental measurements of EWPOs.
Limitations will come from experimental systematics, but they are in many cases estimated to be improved by more than an order of magnitude compared to the LEP experiments \cite{Baer:2013cma,Gomez-Ceballos:2013zzn,dEnterria:2016sca,CEPC-SPPCStudyGroup:2015csa}.
This raises a new situation and theoretical calculations must 
be much more precise than assumed before \cite{mini,poster}. 
The improved precision will provide a platform for deep tests of the quantum structure of nature and unprecedented sensitivity to heavy or super-weakly coupled new physics.

As an important step towards that goal, this article 
reports on the completion of such calculations at the two-loop level in the 
Glashow-\-Weinberg-\-Salam gauge theory, known as the Stan\-dard Mo\-del (SM) 
\cite{Weinberg:1967tq,Glashow:1961tr,Salam:1968rm}.  

\bigskip

The first non-trivial study of electroweak (EW) loop effects was the  calculation of the large quadratic top quark  
mass contribution to the $Z$ 
and $W$ propagators at one-loop order \cite{Veltman:1977kh}. 
A few years later, the on-shell renormalization scheme as it is used today \cite{Sirlin:1980nh} and the notion of 
effective weak mixing angles \cite{Marciano:1980pb} were introduced, and the 
scheme was used for calculations of the $W^{\pm}$ and $Z$ boson masses \cite{Marciano:1983wwa}.
The  complete one-loop corrections to the $Z$ decay parameters were derived in 
Refs.~\cite{Akhundov:1985fc,Beenakker:1988pv,Jegerlehner:1988ak,Bernabeu:1987me}, and those to the $W^{\pm}$ 
width in Refs.~\cite{Bardin:1986fi,Jegerlehner:1988ak,Denner:1990tx}.
Through the years of LEP and SLC studies, the effects of EW corrections became visible in global 
fits of the SM parameters \cite{Hikasa:1992je,ALEPH:2005ablast0,Bardin:1997xq,Bardin:1995XX}. Global fits to EW 
precision measurements allowed to
predict 
the mass of the top quark and the Higgs boson prior to their discoveries
 at Tevatron in 1995  \cite{Abe:1995hr,D0:1995jca}
and
at the LHC in 2012 \cite{Aad:2012tfa}.

At future $e^+e^-$ colliders, EWPOs will again play a crucial role. 
% Here a crucial role plays electroweak precision observables (EWPOs). %play a crucial role in testing the
%Standard Model (SM) at the quantum level and constraining physics beyond the SM.
These include the total and partial widths
of the $Z$ boson and the $Z$-boson couplings. 
The latter can be extracted from
measurements of the cross-section and polarization and angular asymmetries of
the processes $e^+e^- \to (Z) \to f\bar{f}$. Here $f$ stands for any SM lepton or quark, except
the top quark, whereas the notation $(Z)$ is supposed to indicate that the amplitude is dominated by the $s$-channel $Z$-boson resonance, but there is contamination from photon and two-boson backgrounds.

Already for the precision achieved at LEP and SLC, the calculation of loop corrections beyond the one-loop order was necessary to keep theory uncertainties under control. Specifically, these included two-loop $\OO(\alpha\as)$ \cite{Djouadi:1987gn,Djouadi:1987di,Kniehl:1989yc,Kniehl:1991gu,Djouadi:1993ss} and fermionic $\OO(\alpha^2)$ \cite{Barbieri:1992nz,Barbieri:1992dq,Fleischer:1993ub,Fleischer:1994cb,Degrassi:1996mg,Degrassi:1996ps,Degrassi:1999jd,Freitas:2000gg,Freitas:2002ja,Awramik:2004ge,Hollik:2005va,Awramik:2008gi,Freitas:2012sy,Freitas:2013dpa,Freitas:2014hra} corrections to the Fermi constant, which can be used to predict the $W$-boson mass, and to the $Z$-pole parameters.
Here $\alpha$ refers an electroweak loop order, whereas ``fermionic'' denotes contributions from diagrams with at least one closed fermion loop.
In addition, leading three- and four-loop results, enhanced by powers of the top Yukawa coupling $y_\Pt$, were obtained at order $\OO(\at\as^2)$
\cite{Avdeev:1994db,Chetyrkin:1995ix},
$\OO(\at^2\as)$, $\OO(\at^3)$ \cite{vanderBij:2000cg,Faisst:2003px},
and
$\OO(\at\as^3)$ \cite{Schroder:2005db,Chetyrkin:2006bj,Boughezal:2006xk},
where $\at = y_\Pt^2/(4\pi)$.

For the EW two-loop corrections, the calculation of the fer\-mi\-onic contributions was a natural first step, since these are numerically enhanced by the numbers of flavors and colors and by powers of $y_\Pt$. Moreover, the fermionic two-loop diagrams are relatively simpler than the full set. For example, the latter includes non-planar vertex topologies, which are absent in the former. The remaining bosonic two-loop corrections to the Fermi constant and the leptonic
effective weak mixing angle, $\seff{\ell}$, have subsequently been presented in Refs.~\cite{Awramik:2002wn,
Awramik:2003ee,Onishchenko:2002ve,Awramik:2003rn,Awramik:2006ar,Hollik:2006ma,Awramik:2006uz}, and more recently also for the weak mixing angle in the $b\bar{b}$ channel \cite{Dubovyk:2016aqv}. 

While the numerical effects of the bosonic two-loop corrections are relatively small compared to the current experimental precision from LEP and SLC, their inclusion will become mandatory for future $e^+e^-$ colliders. Thus the computation of the full two-loop corrections for all $Z$-pole EWPOs is an important goal.
This article completes this goal by presenting the remaining  bosonic $\OO(\alpha^2)$ contributions to
the $Z$-boson total and partial widths, and the hadronic $Z$-peak cross-section within the SM. This has been achieved by using the numerical integration methods discussed in Ref.~\cite{Dubovyk:2016aqv}, with some technical improvements.
 
The paper is organized as follows. After a brief review of the field theoretic definition of the relevant observables in section~\ref{sc:def}, the technical aspects of the two-loop calculation are described in section~\ref{sc:calc}. The numerical impact of the bosonic EW two-loop corrections is demonstrated in section~\ref{sc:res}. In particular, results for the total and partial $Z$ widths, several commonly used branching ratios, and the hadronic $Z$-peak cross-section are given in terms of simple parameterization formulae, which provide an accurate description of the full results within the currently allowed ranges of the input parameters.
Finally, the theory uncertainty from missing three- and four-loop contributions is estimated in section~\ref{sc:error}, before concluding in section~\ref{sc:summ}.

%%%%%%%%%%%%%%%%%%%%%%%%%%%%%%%%%%%%%%%%%%%%%%%%%%%%%%%%%%%%%%

\section{Definition of the observables}
\label{sc:def}

The amplitude for $e^+e^- \to f \bar{f}$ near the $Z$ pole, $\sqrt{s} \approx
\MZ$ can be written in a theoretically well-defined way 
as a Laurent expansion around the complex pole $s_0 \equiv \mz^2 - i\mz \gz$,
\begin{equation}
{\cal A}[e^+e^- \to f \bar{f}] = \frac{R}{s-s_0} + S +
        (s-s_0) S' +
\dots, \label{polexp}
\end{equation}
where $\mz$ and $\gz$ are the on-shell mass and width of the $Z$ boson,
respectively. According to eq.~\eqref{polexp}, the approximate line shape of the
cross-section near the $Z$ pole is given by $\sigma \propto [(s-\mz^2)^2 +
\mz^2\gz^2]^{-1}$. It is important to note that this differs from the line shape
used in experimental analyses, which is of the form $\sigma \propto [(s-\MZ^2)^2
+ s^2\GZ^2/\MZ^2]^{-1}$. As a result, the parameters in eq.~\eqref{polexp}
differ from the experimental mass $\MZ$ and width $\GZ$ from LEP 
 by a fixed
factor \cite{Bardin:1988xt}:
\begin{align}
\textstyle
\mz &= \MZ\big/\sqrt{1+\Gamma_\PZ^2/\MZ^2}\,, \notag \\
\gz &= \Gamma_\PZ\big/\sqrt{1+\Gamma_\PZ^2/\MZ^2}\,. \label{massrel}
\end{align}
Numerically, this leads to $\mz \approx \MZ - 34\mev$ and $\gz \approx
\Gamma_\PZ - 0.9\mev$.

\vspace{\bigskipamount}
The total width, $\gz$, can be extracted from the condition that the $Z$
propagator has a pole at $s=s_0$, leading to
\begin{equation}
\gz = \frac{1}{\mz} \text{Im}\,\Sigma_\PZ(s_0),
\end{equation}
where $\Sigma_\PZ(s)$ is the transverse part of the $Z$ self-energy. Using the
optical theorem, it can also be written as \cite{Freitas:2013dpa,Freitas:2014hra}
\begin{align}
\gz &= \sum_f \overline{\Gamma}_f, \\ 
\overline{\Gamma}_f &= \frac{N_c^f\mz}{12\pi} \Bigl [
 {\cal R}_{\rm V}^f F_{\rm V}^f + {\cal R}_{\rm A}^f F_{\rm A}^f \Bigr ]_{s=\mz^2} 
 \;, \label{Gz}
\end{align}
Here the sum runs over all fermion types besides the top quark,  $f=e,\mu,\tau,\nu_e,\nu_\mu,\nu_\tau,u,d,c,s,b$,
and $N_c^f = 3(1)$ for quarks (leptons). The radiator functions ${\cal R}_{\rm
V,A}$ capture the effect of final-state QED and QCD corrections. They are known 
up to ${\cal O}(\as^4)$ and $\OO(\alpha^2)$ for massless external fermions
and ${\cal O}(\as^3)$ for the kinematic mass corrections \cite{Chetyrkin:1994js,Baikov:2008jh,Kataev:1992dg}.
For the results shown in this article, the
explicit form given in the appendix of Ref.~\cite{Freitas:2014hra} has been used.

The remaining radiative corrections are IR finite and contained in the
form factors $F_{\rm V,A}^f$. These include massive EW corrections as well as mixed EW--QCD
and EW--QED corrections. The bosonic two-loop contributions, which are of interest
for this article, contribute according to \cite{Freitas:2014hra}:
\begin{align}
F_{\rm V(2)}^f =\;& 2 \,\text{Re}\, (v_{f(0)}v_{f(2)})  + |v_{f(1)}|^2 \nonumber \\&- v_{f(0)}^2
 \bigl [\text{Re}\,\Sigma'_{\PZ(2)}
  - (\text{Re}\,\Sigma'_{\PZ(1)})^2 \bigr ] 
\nonumber \\
& -~2 \,\text{Re}\, (v_{f(0)}v_{f(1)})\;\text{Re}\,\Sigma'_{\PZ(1)} 
 \,, \label{Fv} \\[1ex]
F_{\rm A(2)}^f =\;& 2 \,\text{Re}\, (a_{f(0)}a_{f(2)})  + |a_{f(1)}|^2 \nonumber \\&- a_{f(0)}^2
 \bigl [\text{Re}\,\Sigma'_{\PZ(2)}
  - (\text{Re}\,\Sigma'_{\PZ(1)})^2 \bigr ] 
 \nonumber \\&- 2 \,\text{Re}\, (a_{f(0)}a_{f(1)})\;\text{Re}\,\Sigma'_{\PZ(1)} 
 \,, \label{Fa}
\end{align}
where $v_f$ and $a_f$ are the effective vector and axial-vector couplings,
respectively, which include $Zf\bar{f}$ vertex corrections
and  $Z$--$\gamma$ mixing contributions. $\Sigma'_\PZ$ denotes the derivative of
$\Sigma_\PZ$, and the loop order is indicated by the subscript $(n)$.

It should be pointed out that $v_f$, $a_f$ and $\Sigma_Z$ as defined above include  $\gamma$--$Z$
mixing contributions, $i.\,e.$
\begin{align}
v_f(s) &= v_f^\PZ(s) - v_f^\gamma(s)\, 
 \frac{\Sigma_{\rm \gamma Z}(s)}{s+\Sigma_{\gamma\gamma}(s)}\,, \\
a_f(s) &= a_f^\PZ(s) - a_f^\gamma(s)\, 
 \frac{\Sigma_{\rm \gamma Z}(s)}{s+\Sigma_{\gamma\gamma}(s)}\,, \\
\Sigma_\PZ(s) &= \Sigma_{\rm ZZ}(s) - \frac{[\Sigma_{\rm \gamma
Z}(s)]^2}{s+\Sigma_{\gamma\gamma}(s)}\,.
\end{align}
Here $v_f^\PZ$ and $a_f^\PZ$ are the one-particle irreducible $Zf\bar{f}$
vector and axial-vector vertex contributions, respectively, whereas $v_f^\gamma$
and $a_f^\gamma$ are their counterpart for the $\gamma f\bar{f}$ vertex.
Furthermore, $\Sigma_{V_1V_2}$ denotes the one-particle irreducible $V_1$--$V_2$
self-energy.

Another important quantity is the 
hadronic peak cross section, $\sigma^0_{\rm had}$, which is 
defined as the total cross section for $e^+e^- \to (Z) \to \text{hadrons}$ for
$s=\MZ^2$, after removal of $s$-channel photon exchange and box diagram 
contributions, as well as after the
de-convolution of initial-state and initial-final interference QED effects 
\cite{Bardin:1997xq,ALEPH:2005ablast0}.
The impact of the bosonic two-loop vertex corrections on $\sigma^0_{\rm had}$
is given by \cite{Freitas:2013dpa,Freitas:2014hra}
\begin{align}
\sigma^0_{\rm had(2)} = \sum_{f=u,d,c,s,b}
 \frac{12\pi}{\mz^2} \Biggl[
 &\frac{\overline{\Gamma}_{e(0)}\overline{\Gamma}_{f(2)} +
 \overline{\Gamma}_{e(2)}\overline{\Gamma}_{f(0)}}{{\gz^2}_{(0)}}  \nonumber \\&
- 2\frac{\overline{\Gamma}_{e(0)}\overline{\Gamma}_{f(0)}}{{\gz^2}_{(0)}}
  {\gz^2}_{(2)} \Biggr]. \label{s0had}
\end{align} 
The form factors $F^f_{\rm V,A}$ are understood to include appropriate
counterterms such that they are UV finite. Throughout this work, the on-shell
renormalization scheme is being used, which defines all particle masses in terms
of their (complex) propagator poles and the electromagnetic coupling in terms of the
photon-electron vertex in the Thomson limit. A more detailed discussion of the
relevant counterterms can be found in Ref.~\cite{Freitas:2002ja}.

As a consequence of this renormalization scheme, the EW corrections are organized as a series in the electromagnetic 
coupling $\alpha$, 
rather than the Fermi constant $G_\mu$. Instead, $G_\mu$ will be used to compute $\MW$ within the SM, including 
appropriate two-loop and partial higher-loop corrections. After this step, the remaining input parameters for the 
prediction of the $Z$ coupling form factors are $\MZ$,
$\MH$, $\mt$, $G_\mu$, $\alpha$, $\as$ and $\Delta\alpha$. Here $\Delta\alpha$ captures the running of the 
electromagnetic coupling induced by light fermion loops. It is defined through $\alpha(\MZ^2) =
\alpha(0)/(1-\Delta\alpha)$, where $\alpha(q^2)$ is the coupling at scale $q^2$.
The contribution from leptons to $\Delta\alpha$ can be computed perturbatively and is known at the three-loop level 
\cite{Steinhauser:1998rq-new}, $\Delta\alpha_{\rm lept}(\MZ) = 0.0314976$. On the other hand, the quark contribution is 
non-perturbative at low scales and thus is commonly derived from experimental data. For recent evaluations of 
$\Delta\alpha^{(5)}_{\rm had}$, see Refs.~\cite{Davier:2017zfy,Jegerlehner:2017zsb,Keshavarzi:2018mgv}.
As a reference value, $\Delta\alpha^{(5)}_{\rm had} = 0.02750$ is
used in this work.

Additionally, $\GZ$ and $\GW$ are needed as inputs to convert $\MZ$ and $\MW$ to the complex pole scheme, see 
eq.~\eqref{massrel}. Furthermore, the radiator functions ${\cal R}^f_{\rm V,A}$ depend on $\mb^{\overline{\rm MS}}$, 
$m_{\rm c}^{\overline{\rm MS}}$ and $m_\tau$ to account for kinematic fermion mass effects in the final state, whereas 
the masses of electron, muon, neutrinos, and $u/d/s$ quarks can be taken as zero to very good approximation. In contrast 
to all other masses in this work, the $\overline{\rm MS}$ masses are used for the bottom and charm quarks, since their 
on-shell counterparts are poorly defined.

%%%%%%%%%%%%%%%%%%%%%%%%%%%%%%%%%%%%%%%%%%%%%%%%%%%%%%%%%%%%%%

\section{Calculation of two-loop vertex corrections}
\label{sc:calc}

\label{sec:2l}

For the calculations we followed the strategy developed in Ref.~\cite{Dubovyk:2016aqv}, where the two-loop bosonic corrections 
 to the bottom quark weak mixing angle, $\seff{b}$, were obtained. In fact, the $Zb\bar{b}$ vertex is the technically most 
difficult case due to the larger number 
of mass scales in that problem compared to other flavors. 
Details are described there and also in \cite{Dubovyk:2016ocz,Dubovyk:2016zok,Dubovyk:2017cqw}. 
On the other hand, for the computation of the $Z$ width we are faced not only with ratios $v_{f(2)}/a_{f(2)}$, but also with sums of powers of $v_{f(2)}$ and $a_{f(2)}$, see 
\eqref{Fv} and \eqref{Fa}. 
This leads to the occurrence of extra integrals which cancel out in the ratios $v/a$.

The complete set of two-loop diagrams required for this calculation have been generated with the computer algebra 
package {\tt FeynArts 3.3} \cite{Hahn:2000kx}. They can be divided into several categories. 
The renormalization counterterms require two-loop self-energies with Minkowskian external momenta, 
$p^2 = M_i^2 + i\varepsilon$, $M_i = 
\MW,\, \MZ$. In addition, there are two-loop vertex integrals with
one non-vanishing external momentum squared, $s = \MZ^2 + i\varepsilon$.
The two-loop self-energy integrals needed for the renormalization procedure
 and the vertex integrals with self-energy sub-loops have been computed using the dispersion relation technique 
described in 
Refs.~\cite{Bauberger:1994by,Bauberger:1994hx,Awramik:2006uz}.
The remaining bosonic two-loop diagrams amount to
about one thousand integrals with a planar or non-planar vertex topology.
 
We did not try to reduce these integrals to a minimal set of master integrals, except for trivial
cancellations of numerator and denominator terms. 
This means that tensors of rank $R\leq 3$ were calculated directly.
For this purpose, two numerical approaches were used.
Firstly, sector 
decomposition (SD) \cite{Hepp:1966eg} was applied,  with the packages {\tt SecDec} \cite{Binoth:2000ps,Borowka:2015mxa} 
and  {\tt FIESTA 3} \cite{Smirnov:2013eza}.
Secondly, 
Mellin Barnes (MB) representations \cite{Usyukina:1975yg,Smirnov:1999gc,Tausk:1999vh} were derived and evaluated  with 
the {\tt MBsuite}, consisting of software packages available at the {\tt MBtools} 
webpage in the {\tt hepforge} archive \cite{mbtools}: 
{\tt MB} \cite{Czakon:2005rk}, %and MBasymtotics.m 
{\tt MBresolve} \cite{Smirnov:2009up}, 
{\tt AMBRE 1} \cite{Gluza:2007rt},
{\tt barnesroutines} (D.~Kosower) and
{\tt PlanarityTest} \cite{Bielas:2013rja}, {\tt AMBRE 2} \cite{Gluza:2010rn} 
and {\tt AMBRE 3} \cite{Dubovyk:2015yba}, as well as {\tt 
MBsums} 
\cite{Ochman:2015fho}, which are available from the {\tt AMBRE} webpage \cite{Katowice-CAS:2007}. 
The numerical package {\tt MBnumerics} is being developed since 2015 \cite{Usovitsch:201606}. It is of special 
importance for Minkowskian kinematics as encountered here.
For the numerical integrations, {\tt MBsuite} calls the {\tt CUHRE} routine of the {\tt CUBA} library 
\cite{Hahn:2004fe,Hahn:2014fua}.

Some new  classes of integrals compared to the $\seff{b}$ case are met.
They are simpler from a numerical point of view than those solved in Ref.~\cite{Dubovyk:2016aqv}. For instance, 
there are various one- and 
two-scale integrals with internal $W$ propagators, which improves the singular threshold behaviour of integrals with only $Z$ propagators. 
There are altogether about one hundred integrals of this kind with different permutations of propagators, 
including the tensor integrals. 
As an example of one of the most difficult cases, the SD method for integrals from Fig.~1 in \cite{Dubovyk:2016aqv}
gives an accuracy of up to four relevant digits. Using the MB method, these diagrams are equivalent to up to 
4-dimensional MB integrals, which can be calculated efficiently with eight relevant digits by 
{\tt MBnumerics}.

In select cases, like those described above, the MB approach is uniquely powerful. This statement applies to several hundred integrals. 
In the majority of integrals, though, the SD method is presently more efficient 
than the MB approach, mainly due to the smaller number of 
integration variables. 
For our semi-automatized calculation of massive 2-loop vertices the availability of two complementary numerical methods 
with a large overlap was crucial.

%%%%%%%%%%%%%%%%%%%%%%%%%%%%%%%%%%%%%%%%%%%%%%%%%%%%%%%%%%%%%%

\section{Numerical results}
\label{sc:res}

In this section, numerical results for bosonic two-loop corrections are compared to and combined with all other known corrections to the $Zf\bar{f}$ vertices.
These are
\begin{itemize}
\item Complete one-loop EW contributions \cite{Akhundov:1985fc} (which have been re-evaluated for this work) and fermionic $\OO(\alpha^2)$ contributions \cite{Freitas:2013dpa,Freitas:2014hra}; 
\item Mixed QCD-EW corrections to internal gauge-boson self-energies of order $\OO(\alpha\as)$ 
\cite{Djouadi:1987gn,Djouadi:1987di,Kniehl:1989yc,Kniehl:1991gu,Djouadi:1993ss}
(where again we use our own re-evaluation of these terms);
\item Higher-loop corrections in the large-$\mt$ limit, of
order $\OO(\at\as^2)$ \cite{Avdeev:1994db,Chetyrkin:1995ix}, $\OO(\at^2\as)$, $\OO(\at^3)$ 
\cite{vanderBij:2000cg,Faisst:2003px},
and $\OO(\at\as^3)$ \cite{Schroder:2005db,Chetyrkin:2006bj,Boughezal:2006xk},
where $\at \equiv y_\Pt^2/(4\pi)$ and $y_\Pt$ is the top Yukawa coupling;
\item Final-state QED radiation and, for quark final states, QCD radiation up to
$\OO(\alpha^2)$, $\OO(\alpha\as)$ and $\OO(\as^4)$ 
\cite{Chetyrkin:1994js,Baikov:2008jh,Kataev:1992dg};
incorporated through the radiator functions ${\cal R}_{\rm V,A}$ in \eqref{Fv} and \eqref{Fa};
\item Non-factorizable $\OO(\alpha\as)$ vertex contributions
\cite{Czarnecki:1996ei,Harlander:1997zb,Fleischer:1992fq,Buchalla:1992zm,Degrassi:1993ij,Chetyrkin:1993jp},
 %\cite{nfact,nfactb}, 
which cannot be written as a product of 
EW form factors $F_{\rm V,A}$ and final-state radiator functions ${\cal
R}_{\rm V,A}$, but instead are added separately to the formula in \eqref{Gz}.
\end{itemize}
These are applied to a range of EWPOs: The partial $Z$ widths, $\Gamma_f \equiv
\Gamma(Z \to f\bar{f})$, as well as total width, $\GZ$, various branching ratios, and the hadronic peak 
cross-section $\sigma_{\rm had}$. 
The full electroweak two-loop corrections for the leptonic and bottom-quark
asymmetries have been published previously \cite{Awramik:2006ar,Hollik:2006ma,Awramik:2006uz,Dubovyk:2016aqv} 
and are not repeated here. 
Nevertheless, as a cross-check we reproduced the result for the leptonic asymmetry and 
found agreement with Refs.~\cite{Awramik:2006ar,Hollik:2006ma} within intrinsic numerical uncertainties. 
Moreover, with the methods described here we can produce results for the bosonic two-loop corrections to $\seff{\ell}$ 
with four robust digits of precision, which exceeds the accuracy obtained with asymptotic expansions as in 
Ref.~\cite{Awramik:2006ar}.

As discussed above, the gauge-boson mass renormalization has been performed in accordance with the 
complex-pole scheme in eq.~\eqref{polexp}. However, for the sake of comparison with the wider literature, the numerical 
results below are 
presented after translating to the scheme with an $s$-dependent width. In other words, results are shown for un-barred 
quantities, such as \GZ\ in eq.~\eqref{massrel}.

Light fermion masses $\mf$, $f\neq t$, have been neglected throughout, except for a non-zero bottom quark mass in the $\OO(\alpha)$ and
$\OO(\alpha\as)$ vertex contributions, as well as for non-zero $\mb$, $m_{\rm c}$ and $m_\tau$ in the
radiators ${\cal R}_{\rm V,A}$.  
  The numerical input values used in this section are listed in
Tab.~\ref{tab:input}.

\begin{table}[tb]
\renewcommand{\arraystretch}{1.2}
\begin{center}
\begin{tabular}{|ll|}
\hline
Parameter & Value   \\
\hline \hline
$\MZ$ & 91.1876 GeV  \\
$\Gamma_\PZ$ & 2.4952 GeV \\
$\MW$ & 80.385 GeV \\
$\Gamma_\PW$ & 2.085 GeV \\
$\MH$ & 125.1 GeV  \\
$\mt$ & 173.2 GeV  \\
 $\mb^{\overline{\rm MS}}$ & 4.20 GeV \\
 $m_{\rm c}^{\overline{\rm MS}}$ & 1.275 GeV \\
 $m_\tau$ & 1.777 GeV \\
  $m_e,m_\mu,m_u,m_d,m_s$ & 0 \\
 $\Delta\alpha$ & 0.05900 \\
 $\as(\MZ)$ & 0.1184 \\
 $G_\mu$ & $1.16638 \times 10^{-5}$~GeV$^{-2}$ \\
\hline
\end{tabular}
\end{center}
\vspace{-2ex}
\mycaption{Input parameters used in the numerical analysis,
from \cite{Patrignani:2016xqp}, except for $\Delta\alpha$, for which a value close to several recent evaluations \cite{Davier:2017zfy,Jegerlehner:2017zsb,Keshavarzi:2018mgv} has been chosen.
\label{tab:input}}
\end{table}
%-------------------------------------------------------------

\subsection{Partial widths}

\begin{center}
%-------------------------------------------------------------
\begin{table*}[tbp]
\renewcommand{\arraystretch}{1.2}
\begin{center}
\begin{tabular}{|l|r|r|r|r|r|r|}
\hline
\multicolumn{1}{|r|}{$\Gamma_i$ [MeV]} & $\Gamma_e\;\;$ & $\Gamma_\nu\;\;$ & $\Gamma_d\;\;$ & $\Gamma_u\;\;$ & 
 $\Gamma_b\;\;$ & $\Gamma_\PZ\;\;$ \\
\hline \hline
Born & 81.142 & 160.096 & 371.141 & 292.445 & 369.562 & 2420.19
\\
$\OO(\alpha)$ & 2.273 & 6.174 & 9.717 & 5.799 & 3.857 & 60.22 
\\
$\OO(\alpha\as)$ & 0.288 & 0.458 & 1.276 & 1.156 & 2.006 & 9.11 
\\
$\OO(\alpha_\Pt\as^2,\,\alpha_\Pt\as^3,\,\alpha^2_\Pt\as,\,\alpha_\Pt^3)$ &
 0.038 & 0.059 & 0.191 & 0.170 & 0.190 & 1.20 \\
$\OO(N_f^2\alpha^2)$ & 0.244 & 0.416 & 0.698 & 0.528 & 0.694 & 5.13 
\\
$\OO(N_f\alpha^2)$ & 0.120 & 0.185 & 0.493 & 0.494 & 0.144 & 3.04 \\
$\OO(\alpha^2_{\rm bos})$ & 
                    0.017 & 0.019 & 0.059 & 0.058 & 0.167 & 0.51 \\
\hline
\end{tabular}
\end{center}
%\vspace{-2ex}
\mycaption{Contributions of different orders in perturbation theory to the 
partial and total $Z$ widths. A fixed value of $\MW$ has been used as input, 
instead of $G_\mu$. $N_f$ and $N_f^2$ refer to corrections with 
one and two closed fermion loops, respectively, whereas $\alpha^2_{\rm bos}$
denotes contributions without closed fermion loops. Furthermore, $\alpha_\Pt = y_\Pt^2/(4\pi)$.
In all rows the radiator functions ${\cal R}_{\rm V,A}$ with known contributions
through $\OO(\as^4)$, $\OO(\alpha^2)$ and $\OO(\alpha\as)$ are included.
\label{tab:res1}}
\end{table*}
\end{center}
%-------------------------------------------------------------
 
Let us begin by presenting results for a fixed value of $\MW$ as input, instead of calculating $\MW$ from $G_\mu$. 
This more clearly illustrates the impact of the newly completed $\OO(\alpha_{\rm bos})^2$ corrections.
Table~\ref{tab:res1} shows the contributions from different loop orders to the SM prediction of various partial $Z$ widths. 
As is evident from the table, the two-loop
EW corrections are significant and larger than the current experimental uncertainty (2.3 GeV for $\Gamma_\PZ$ 
\cite{ALEPH:2005ablast0}).
The newly calculated bosonic corrections $\OO(\alpha^2_{\rm bos})$ are smaller but still noteworthy. They
amount to half of all known leading three-loop 
QCD corrections
$\OO(\alpha_\Pt\as^2$, $\alpha_\Pt\as^3$, $\alpha^2_\Pt\as$, $\alpha_\Pt^3)$, even though the latter are enhanced by 
powers of $\as$, $\at$ and $N_f$.

Table~\ref{tab:res2} shows the SM predictions obtained if one uses $G_\mu$ as an input to compute $\MW$, based on
the results of
\cite{Freitas:2000gg,Awramik:2002wn,Awramik:2003ee,Onishchenko:2002ve,Freitas:2002ja,Awramik:2003rn}. Each line of 
the table adds an additional order of perturbation theory to the previous line, using the same order for the $Zf\bar{f}$ 
vertex corrections and the calculation of the $W$ mass\footnote{Note that the value in the next-to-last line of 
Tab.~\ref{tab:res2} differs slightly from Ref.~\cite{Freitas:2013dpa}. This is because in Ref.~\cite{Freitas:2013dpa} 
the ``best value'' prediction of $\MW$ was carried with the full (fermionic plus bosonic) EW two-loop predictions 
included. Here, however, we are interested in a clear distinction of fermionic and bosonic two-loop terms in all 
contributions, including the $\MW$ prediction.}.

The $\OO(\alpha^2_{\rm bos})$ correction to $\Gamma_\PZ$, corresponding to the difference between the last two rows in Table~\ref{tab:res2}, amounts to 0.34~MeV, which is more than three times 
larger than its previous estimation \cite{Freitas:2014hra}. An updated discussion on how this knowledge changes the 
intrinsic error estimations will be given in section~\ref{sc:error}. 

%-------------------------------------------------------------
\begin{table}[tb]
\renewcommand{\arraystretch}{1.2}
\begin{center}
\begin{tabular}{|l|r|r|}
\hline
 & $\Gamma_\PZ$ [GeV] & $\sigma^0_{\rm had}$ [nb] \\
\hline \hline
Born & 2.53601 & 41.6171 \\
$+~\OO(\alpha)$ & 2.49770 & 41.4687 \\
$+~\OO(\alpha\as)$ & 2.49649 & 41.4758 \\
$+~\OO(\alpha_\Pt\as^2,\,\alpha_\Pt\as^3,\,\alpha^2_\Pt\as,\,\alpha_\Pt^3)$ &
% 2.49594 & 41.4769 \\
2.49560 & 41.4770 \\
$+~\OO(N_f^2\alpha^2,N_f\alpha^2)$ & 2.49441 & 41.4883 \\
$+~\OO(\alpha^2_{\rm bos})$ & 2.49475 & 41.4896 \\
\hline
\end{tabular}
\end{center}
\vspace{-2ex}
\mycaption{Results for $\Gamma_\PZ$ and $\sigma^0_{\rm had}$, with $\MW$ calculated
from $G_\mu$ using the same order of perturbation theory as indicated in each
line. In all cases, the complete radiator functions ${\cal R}_{\rm V,A}$
are included.
\label{tab:res2}}
\end{table}
%-------------------------------------------------------------

\bigskip
\subsection{Ratios}

The experimental results from LEP and SLC are typically not presented in terms of partial widths for the different final 
states. 
Instead, this information is captured in the form of various branching ratios. The most relevant ones are
\begin{align}
\hspace{-1em}
R_\ell &\equiv \Gamma_{\rm had}/\Gamma_\ell, &
R_c &\equiv  \Gamma_c/\Gamma_{\rm had}, &
R_b &\equiv  \Gamma_b/\Gamma_{\rm had}, \label{rat}
\end{align}
where $\Gamma_\ell = \frac{1}{3}(\Gamma_e + \Gamma_\mu + \Gamma_\tau)$, and
$\Gamma_{\rm had}$ is the partial width into hadronic final states, which at the parton level is
equivalent to $\sum_q \Gamma_q$ ($q=u,d,c,s,b$).

In addition, the hadronic peak cross-section \eqref{s0had} is, to a good approximation, defined as a ratio of partial widths and the total $Z$ width.

%-------------------------------------------------------------
\begin{table*}[tb]
\renewcommand{\arraystretch}{1.2}
\begin{center}
\begin{tabular}{|l|c|c|c|}
\hline
 & $R_\ell$ & $R_c$ & $R_b$ \\
\hline \hline
Born & 21.0272 & 0.17306 & 0.21733 \\
$+~\OO(\alpha)$ & 20.8031 & 0.17230 & 0.21558 \\
$+~\OO(\alpha\as)$ & 20.7963 & 0.17222 & 0.21593 \\
$+~\OO(\alpha_\Pt\as^2,\,\alpha_\Pt\as^3,\,\alpha^2_\Pt\as,\,\alpha_\Pt^3)$ &
20.7943 & 0.17222 & 0.21593 \\
$+~\OO(N_f^2\alpha^2,N_f\alpha^2)$ & 20.7512 & 0.17223  & 0.21580 \\
$+~\OO(\alpha^2_{\rm bos})$ & 20.7516 & 0.17222 & 0.21585 \\
\hline
\end{tabular}
\end{center}
\vspace{-2ex}
\mycaption{Results for the ratios $R_\ell$, $R_c$ and $R_b$, with $\MW$ calculated
from $G_\mu$ to the same order as indicated in each
line. 
In all cases, the complete radiator functions ${\cal R}_{\rm V,A}$
are included. 
\label{tab:resr}}
\end{table*}
%-------------------------------------------------------------

Numerical results for $\sigma^0_{\rm had}$ and the ratios in \eqref{rat} are given in Tab.~\ref{tab:res2} and 
Tab.~\ref{tab:resr}, respectively, again broken down to different orders of radiative corrections. These quantities are 
less sensitive to higher loop effects than $\Gamma_\PZ$, 
since there is a partial cancellation between the corrections in the numerators and denominators of the ratios.
Thus the influence of the new bosonic corrections on all branching ratios $R_\ell,R_c,R_b$ and on $\sigma^0_{\rm had}$ 
is about 0.02\% or less, which 
is far below the current experimental errors: $R_\ell = 20.767 \pm 0.025$, $R_c = 0.1721 \pm
0.0030$, $R_b = 0.21629 \pm 0.00066$, and $\sigma^0_{\rm had} = 41.541 \pm 0.037$~nb \cite{ALEPH:2005ablast0}. 
However, these are at the level of sensitivity of proposed measurements of $R_b$ at future $e^+e^-$ colliders 
\cite{Baer:2013cma,Gomez-Ceballos:2013zzn,dEnterria:2016sca,CEPC-SPPCStudyGroup:2015csa}

\subsection{Parameterization formulae}
\label{sc:fit1}

%\onecolumn
%-------------------------------------------------------------
%-------------------------------------------------------------
\begin{table*}[tb]
\renewcommand{\arraystretch}{1.3}
\begin{center}
\begin{tabular}{|l|cccccccc|c|}
\hline
Observable & $X_0$ & $c_1$ & $c_2$ & $c_3$ & $c_4$ & $c_5$ & $c_6$ & $c_7$ &
max.\ dev. \\
\hline
$\Gamma_{e,\mu}$ [MeV] &
 83.983 & $-$0.061 & 0.810 & $-$0.096 & $-$0.01 & 0.25 & $-$1.1 & 286
 & $<0.001$\\
$\Gamma_{\tau}$ [MeV] &
 83.793 & $-$0.060 & 0.810 & $-$0.095 & $-$0.01 & 0.25 & $-$1.1 & 285
 & $<0.001$\\
$\Gamma_{\nu}$ [MeV] &
 167.176 & $-$0.071 & 1.26 & $-$0.19  & $-$0.02 & 0.36 & $-$0.1 & 504
 & $<0.001$\\
$\Gamma_{u}$ [MeV] &
 299.993 & $-$0.38 &  4.08 &    14.27 &     1.6 & 1.8  & $-$11.1& 1253
 & $<0.002$\\
$\Gamma_{c}$ [MeV] &
 299.916 & $-$0.38 &  4.08 &    14.27 &     1.6 & 1.8  & $-$11.1& 1253
 & $<0.002$\\
$\Gamma_{d,s}$ [MeV] &
 382.828 & $-$0.39  & 3.83 &    10.20 & $-$2.4  & 0.67 & $-$10.1& 1470
 & $<0.002$\\
$\Gamma_{b}$ [MeV] &
 375.889 & $-$0.36 &$-$2.14&    10.53 & $-$2.4  & 1.2  & $-$10.1& 1459
 & $<0.006$\\
$\GZ$ [MeV] & 
 2494.74 & $-$2.3   & 19.9 &    58.61 & $-$4.0  & 8.0  & $-$56.0& 9273
 & $<0.012$\\
\hline
$R_\ell$ [$10^{-3}$] &
 20751.6 & $-$7.8   & $-$37&   732.3  & $-$44   & 5.5  & $-$358 & 11696
& $<0.1$ \\
$R_c$ [$10^{-3}$] &
 172.22 & $-$0.031  & 1.0  &    2.3   &    1.3  & 0.38 & $-$1.2 & 37
& $<0.01$ \\
$R_b$ [$10^{-3}$] &
 215.85 &  0.029   &$-$2.92& $-$1.32 & $-$0.84 & 0.032 & 0.72   & $-$18
& $<0.01$ \\
\hline
$\sigma^0_{\rm had}$ [pb] & 
 41489.6 &    1.6   & 60.0 & $-$579.6 &   38    & 7.3  &   85   & $\!\!\!\!-$86011
 & $<0.1$\\
\hline
\end{tabular}
\end{center}
\vspace{-2ex}
\mycaption{Coefficients for the parameterization formula \eqref{par1} for various
observables ($X$). Within the ranges $\MH = 125.1\pm 5.0\gev$, $\mt = 173.2\pm 4.0\gev$,
$\as=0.1184\pm 0.0050$, $\Delta\alpha = 0.0590 \pm 0.0005$ and $\MZ = 91.1876 \pm
0.0042 \gev$, the formulae approximate the full results with maximal deviations
given in the last column.
\label{tab:fit1}}
\end{table*}
%-------------------------------------------------------------
% \twocolumn

While the tables above only contain numbers for a single benchmark point, the results for a range of input values can 
be conveniently expressed in terms of simple para\-meteri\-zation formulae. The coefficients of these formulae 
have been fitted to the full calculation results on a grid that spans the currently allowed experimental ranges for 
each input parameter. Here the full calculation includes all higher-order corrections listed at
the beginning of section~\ref{sc:res} for the partial widths, branching ratios
and the peak cross-sections, and with $\MW$ calculated from $G_\mu$ to the
same precision\footnote{Fit formulae for the leptonic and bottom-quark asymmetries can be found in Refs.~\cite{Awramik:2006ar,Awramik:2006uz,Dubovyk:2016aqv}.}. For all EWPOs reported here, the same form of parameterization formula is utilized:
\begin{align}
X = X_0 &+ c_1 L_\PH + c_2 \Delta_\Pt + c_3 \Delta_{\as} + c_4 \Delta_{\as}^2
 \nonumber \\ &+ c_5 \Delta_{\as}\Delta_\Pt 
 + c_6 \Delta_\alpha + c_7 \Delta_\PZ, \label{par1} 
 \end{align}
\begin{align}
&L_\PH = \log\frac{\MH}{125.7\gev}, &
 \Delta_\Pt &= \Bigl (\frac{\mt}{173.2\gev}\Bigr )^2-1, \quad \nonumber \\
& \Delta_{\as} = \frac{\as(\MZ)}{0.1184}-1, & \Delta_\alpha &= \frac{\Delta\alpha}{0.059}-1,  \nonumber \\
& \Delta_\PZ = \frac{\MZ}{91.1876\gev}-1. \nonumber
\end{align}
As before, $\MH$, $\MZ$, $\mt$ and $\Delta\alpha$ are defined in the on-shell
scheme, using the $s$-dependent width scheme for $\MZ$ (to match the published
experimental values), while $\as$ is defined in the $\overline{\rm MS}$ scheme. The dependence on $\mb$, $m_{\rm c}$ and 
$m_\tau$ is 
negligible within the allowed ranges for these quantities.

The fit values of the coefficients for the different EWPOs are given in Tab.~\ref{tab:fit1}. With these parameters, 
the formulae
provide very good approximations to the full results within the ranges $\MH =
125.1\pm 5.0\gev$, $\mt = 173.2\pm 4.0\gev$, $\as=0.1184\pm 0.0050$,
$\Delta\alpha = 0.0590 \pm 0.0005$ and $\MZ = 91.1876 \pm 0.0042 \gev$, with
maximal deviations as quoted in the last column of Tab.~\ref{tab:fit1}.
As can be seen from the latter, the accuracies of the fit formulae are sufficient for the forseeable future.

%%%%%%%%%%%%%%%%%%%%%%%%%%%%%%%%%%%%%%%%%%%%%%%%%%%%%%%%%%%%%%

\section{Error estimates}
\label{sc:error}

In addition to the dependence on the input parameters, the accuracy of the
results presented here is limited by unknown three- and four-loop contributions.
The numerically leading missing pieces are the $\OO(\alpha^3)$, $\OO(\alpha^2\as)$,
$\OO(\alpha\as^2)$ and $\OO(\alpha\as^3)$ corrections beyond the known leading 
$y_\Pt^n$ terms from Refs.~\cite{Avdeev:1994db,Chetyrkin:1995ix,vanderBij:2000cg,Faisst:2003px,Schroder:2005db,Chetyrkin:2006bj,Boughezal:2006xk}.

Following Refs.~\cite{Freitas:2014hra,Freitas:2016sty}, the size of these terms may be
estimated by assuming that the perturbation series
approximately is a geometric series. In this way one obtains
\begin{equation}
\begin{aligned}
\OO(\alpha^3)-\OO(\alpha_\Pt^3) &\sim 
 \frac{\OO(\alpha^2)-\OO(\alpha_\Pt^2)}{\OO(\alpha)}
 \OO(\alpha^2), \\
\OO(\alpha^2\as)-\OO(\alpha_\Pt^2\as) &\sim 
 \frac{\OO(\alpha^2)-\OO(\alpha_\Pt^2)}{\OO(\alpha)}
 \OO(\alpha\as), \\
\OO(\alpha\as^2)-\OO(\alpha_\Pt\as^2) &\sim 
 \frac{\OO(\alpha\as)-\OO(\alpha_\Pt\as)}{\OO(\alpha)}
 \OO(\alpha\as), \\
\OO(\alpha\as^3)-\OO(\alpha_\Pt\as^3) &\sim 
 \frac{\OO(\alpha\as)-\OO(\alpha_\Pt\as)}{\OO(\alpha)}
 \OO(\alpha\as^2),
\end{aligned} \label{errprop}
\end{equation}
where the known leading large-$\mt$ approximations have been subtracted in the
numerators. For the example of the total $Z$ width, these expressions lead to
\begin{equation}
\begin{aligned}
\Gamma_\PZ: \quad 
& \OO(\alpha^3)-\OO(\alpha_\Pt^3) \sim 0.20\mev, \\
&\OO(\alpha^2\as)-\OO(\alpha_\Pt^2\as) \sim 0.21\mev, \\
&\OO(\alpha\as^2)-\OO(\alpha_\Pt\as^2) \sim 0.23\mev, \\
&\OO(\alpha\as^3)-\OO(\alpha_\Pt\as^3) \sim 0.035\mev.
\end{aligned}
\label{errgz}
\end{equation}
An additional source of theoretical uncertainty stems from the unknown
$\OO(\as^5)$ final-state QCD corrections and three-loop mixed QED/QCD
final-state correc\-tions of order $\OO(\alpha\as^2)$ and $\OO(\alpha^2\as)$.
In \cite{Freitas:2014hra}
they were found to be sub-dominant, and the estimates can be
taken over from there without change.
Combining these findings with eqs.~\eqref{errgz} in quadrature, 
the total theory error adds up to $\delta\GZ\approx 0.4\mev$. Compared to the previous theory error estimate $\delta\GZ\approx 0.5\mev$ \cite{Freitas:2014hra}
one observes a slight decrease due to the knowledge of the bosonic corrections calculated in this work.

In addition to the elimination of an uncertainty associated with the previous unknown
${\cal O}(\alpha_{\rm bos}^2)$ corrections, the values in the first and second rows of \eqref{errgz} also shifted since 
the full ${\cal O}(\alpha^2)$ corrections used in \eqref{errprop} were not available before. These shifts conspire to 
result in a reduction of the uncertainty estimate for these two error sources.

%-------------------------------------------------------------
\begin{table}[tbp]
\renewcommand{\arraystretch}{1.2}
\begin{center}
\begin{tabular}{|l|l||l|l||l|l|}
\hline
$\Gamma_{e,\mu\,\tau}$ & 0.018~MeV &
$\Gamma_{u,c}$ & 0.11~MeV &
$R_\ell$ & $6\cdot 10^{-3}$ \\
$\Gamma_\nu$ & 0.016~MeV &
$\Gamma_b$ & 0.18~MeV &
$R_c$ & $5\cdot 10^{-5}$ \\
%\cline{3-4}
$\Gamma_{d,s}$ & 0.08~MeV &
$\GZ$ & 0.4~MeV &
$R_b$ & $1\cdot 10^{-4}$ \\
\hline
\end{tabular}
\end{center}
\vspace{-2ex}
\mycaption{Theory uncertainty estimates 
for the partial and total $Z$ widths and branching ratios from missing 3-loop and higher orders. See text for details.
\label{tab:err}}
\end{table}
%-------------------------------------------------------------

\vspace{\bigskipamount}
The corresponding error estimates for the partial widths are shown in 
Table~\ref{tab:err}. For the ratios ($R_\ell$, $R_c$ and $R_b$), the
theory uncertainty has been estimated from the partial widths 
using simple Gaussian error propagation.

%\vspace{\bigskipamount}
The theory uncertainty for the hadronic peak cross-section is dominated by a non-factorizable contribution stemming from the imaginary part of the $Z$-boson self-energy \cite{Freitas:2014hra}. This non-factorizable term does not receive any bosonic two-loop corrections, so that its error estimate can be taken from Ref.~\cite{Freitas:2014hra} without change:
\begin{equation}
\sigma^0_{\rm had}: \quad 
\OO(\alpha^3) \sim 3.7 \text{ pb} , \qquad 
\OO(\alpha^2\as) \sim 4.2 \text{ pb}.
\label{errsig}
\end{equation}
Adding these in quadrature leads to the overall uncertainty estimate of
$\delta\sigma^0_{\rm had} \approx 6$~pb.

%%%%%%%%%%%%%%%%%%%%%%%%%%%%%%%%%%%%%%%%%%%%%%%%%%%%%%%%%%%%%%

\section{Summary}
\label{sc:summ}

In this work the bosonic two-loop electroweak corrections,  $\OO(\alpha^2_{\rm bos})$, to $Z$ boson production and decay
parameters are presented for the first time. These corrections are comparable in size to the leading three-loop corrections of $\OO(\alpha_\Pt\as^2)$,
$\OO(\alpha_\Pt\as^3)$, $\OO(\alpha^2_\Pt\as)$,
$\OO(\alpha_\Pt^3)$.
This is especially pronounced for $\Gamma_b$, see Tab.~\ref{tab:res1},  and for $\sigma_{had}^0$,
see Tab.~\ref{tab:res2}.
The bosonic corrections shift the value of $\Gamma_Z$ by 0.51 MeV when using $\MW$ as input and 0.34 MeV when using $G_\mu$ are input, which is
large from the point of view of future colliders.
The most ambitious FCC-ee project predicts an accuracy of 
0.1 MeV.  Similarly, the bosonic corrections
are important for $R_b$, see Tab.~\ref{tab:resr}. Due to the high accuracy of the numerical loop integrations,
the results obtained here are stable enough even in the context of potential future experimental precisions.

Updated theory error estimations are given, which are slightly reduced due to the newly available full two-loop corrections.
We expect that the numerical integration methods used here can be extended to compute the full three-loop corrections to 
Z-pole EWPOs. For a more detailed discussion of future projections, see 
Ref.~\cite{mini,poster}. However, this is very demanding and needs 
more effort and 
resources. Further, at this level of complexity independent cross-checks by different groups, using independent 
calculations and approaches, are welcome.

It should be noted that the $\OO(\alpha^2_{\rm bos})$ correction for the total $Z$ decay width appears to be relatively  large compared to previous estimates based on the knowledge of the lower order result $\OO(\alpha_{\rm bos})$.
A similar observation concerns the bosonic two-loop corrections to $A_b$.
This means that all estimations at this level of accuracy should be taken with a grain of salt. Therefore, explicit calculations are important even for contributions that were previously estimated to be subdominant.

At this point we should mention that we did not consider the theoretical efforts needed to unfold the large QED corrections from the
measured real cross sections in the $Z$ peak region and to extract 
the EWPOs
studied here in detail. For LEP, this was based on tools such as the ZFITTER package
\cite{Bardin:1999yd,Arbuzov:2005ma,Akhundov:2013ons} and was discussed carefully $e.\,g.$ in 
Refs.~\cite{Bardin:1997xq,Bardin:1999gt,ALEPH:2005ablast0}.
The correct unfolding framework for extracting $2\to 2$ observables at accuracies amounting to about 1/20 of the LEP era 
certainly has to rely on the correct treatment of Laurent series for 
the $Z$ line shape as is discussed $e.\,g.$ in \cite{Leike:1991pq,Riemann:1992gv,Riemann:2015wpn,QED-3loops}.

The 1-loop corrections  to the $Z$ boson
parameters were determined in the 1980s \cite{Akhundov:1985fc}.
Today, 33 years later, while the present study finalizes the determination of the electroweak two-loop 
corrections to the $Z$-boson
parameters, we are already faced with the need of more precision in the future.

%%%%%%%%%%%%%%%%%%%%%%%%%%%%%%%%%%%%%%%%%%%%%%%%%%%%%%%%%%%%%%

\section*{Acknowledgments}
The work of \textit{I.D.}\ is supported by a research grant of Deutscher Akademischer Austauschdienst (DAAD) and by 
Deutsches Elektronensychrotron DESY.
The work of \textit{A.F.}\  is supported in part by the National Science Foundation under
grant PHY-1519175. 
The work of \textit{J.G.} is supported in part by the Polish National Science Centre under
grant no.\ 2017/25/B/ST2/01987 and COST Action CA16201 PARTICLEFACE.
The work of \textit{T.R.}\ is supported in part by an
Alexander von Humboldt Polish Honorary Research Fellowship. 
\textit{J.U.}\ received funding from the European Research Council (ERC) under the European Union's Horizon 2020 research and innovation programme under grant agreement no.\ 647356 (CutLoops). We would like to thank Peter Uwer and his group ``Phenomenology of Elementary Particle Physics beyond the Standard Model'' at Humboldt-Universit\"at zu Berlin for providing computer resources.

\bigskip

%%%%%%%%%%%%%%%%%%%%%%%%%%%%%%%%%%%%%%%%%%%%%%%%%%%%%%%%%%%%%%

%\appendix
%\section{}

%\bibliographystyle{elsarticle-num} % => long references,  with titles and doi and arxiv
%%%%%%\bibliographystyle{utphys_zbb}     % => short references, no titles or doi or arxiv, needs a bit polishing
%\bibliography{2loops_zbos2}

\begin{thebibliography}{100}
\expandafter\ifx\csname url\endcsname\relax
  \def\url#1{\texttt{#1}}\fi
\expandafter\ifx\csname urlprefix\endcsname\relax\def\urlprefix{URL }\fi
\expandafter\ifx\csname href\endcsname\relax
  \def\href#1#2{#2} \def\path#1{#1}\fi

\bibitem{ALEPH:2005ablast0}
S.~Schael, et~al., {Precision electroweak measurements on the $Z$ resonance},
  Phys. Rept. 427 (2006) 257--454.
\newblock \href {http://arxiv.org/abs/hep-ex/0509008}
  {\path{arXiv:hep-ex/0509008}}, \href
  {http://dx.doi.org/10.1016/j.physrep.2005.12.006}
  {\path{doi:10.1016/j.physrep.2005.12.006}}.

\bibitem{Bardin:1997xq}
D.~Bardin, W.~Beenakker, M.~Bilenky, W.~Hollik, M.~Martinez, G.~Montagna,
  O.~Nicrosini, V.~Novikov, L.~Okun, A.~Olshevsky, G.~Passarino, F.~Piccinini,
  S.~Riemann, T.~Riemann, A.~Rozanov, F.~Teubert, M.~Vysotsky, {Electroweak
  working group report}  7--162, CERN 95--03A.~In \cite{Bardin:1995XX}.
\newblock \href {http://arxiv.org/abs/hep-ph/9709229}
  {\path{arXiv:hep-ph/9709229}}.

\bibitem{Bardin:1995XX}
{D. Bardin, W. Hollik, G. Passarino (eds.)}, {Reports} of the working group on
  precision calculations for the $Z$ resonance, report CERN 95-03 (31 March
  1995), \url{http://cds.cern.ch/record/280836/files/CERN-95-03.pdf?version=2}.

\bibitem{Baer:2013cma}
H.~Baer, T.~Barklow, K.~Fujii, Y.~Gao, A.~Hoang, S.~Kanemura, J.~List, H.~E.
  Logan, A.~Nomerotski, M.~Perelstein, et~al., {The International Linear
  Collider Technical Design Report - Volume 2: Physics}.~ (2013).
\newblock \href {http://arxiv.org/abs/1306.6352} {\path{arXiv:1306.6352}}.

\bibitem{Gomez-Ceballos:2013zzn}
M.~Bicer, et~al., {First Look at the Physics Case of TLEP}, JHEP 01 (2014) 164.
\newblock \href {http://arxiv.org/abs/1308.6176} {\path{arXiv:1308.6176}},
  \href {http://dx.doi.org/10.1007/JHEP01(2014)164}
  {\path{doi:10.1007/JHEP01(2014)164}}.

\bibitem{dEnterria:2016sca}
D.~d'Enterria,
  \href{http://inspirehep.net/record/1421932/files/arXiv:1602.05043.pdf}{{Physics
  at the FCC-ee}}, in: {Proceedings of the $17^{th}$ Lomonosov Conference on
  Elementary Particle Physics, August 20-26, 2015, Moscow, Russia}, 2017, pp.
  182--191.
\newblock \href {http://arxiv.org/abs/1602.05043} {\path{arXiv:1602.05043}},
  \href {http://dx.doi.org/10.1142/9789813224568_0028}
  {\path{doi:10.1142/9789813224568_0028}}.
\newline\urlprefix\url{http://inspirehep.net/record/1421932/files/arXiv:1602.05043.pdf}

\bibitem{CEPC-SPPCStudyGroup:2015csa}
{CEPC-SPPC Study Group}, {CEPC-SPPC Preliminary Conceptual Design Report. 1.
  Physics and Detector, IHEP-CEPC-DR-2015-01, IHEP-TH-2015-01, IHEP-EP-2015-01
  (2015)}, \url{http://inspirehep.net/record/1395734/files/main_preCDR.pdf}.

\bibitem{mini}
A. Blondel, J. Gluza, P. Janot (org.), Mini workshop on precision EW and QCD
  calculations for the FCC studies: methods and techniques, 12-13 Jan 2018,
  CERN. Webpage \url{https://indico.cern.ch/event/669224/}. J. Gluza, S. Jadach
  and T. Riemann (eds.), Report to appear.

\bibitem{poster}
I.~Dubovyk, A.~ Freitas, J.~ Gluza, K. Grzanka, S. ~Jadach, T~Riemann and J.~
  Usovitsch, Precision calculations for the $Z$ line shape at the FCC-ee.
  Poster presented at the FCC Week 2018, 9-13 April 2018, Amsterdam, \url{  
https://indico.cern.ch/event/656491/contributions/2947663/attachments/1622685/2582801/Poster-FCC-Amsterdam_SJadach_et_al
  .pdf}.
  
\bibitem{Weinberg:1967tq}
S.~Weinberg, {A Model of Leptons}, Phys. Rev. Lett. 19 (1967) 1264--1266.
\newblock \href {http://dx.doi.org/10.1103/PhysRevLett.19.1264}
  {\path{doi:10.1103/PhysRevLett.19.1264}}.

\bibitem{Glashow:1961tr}
S.~Glashow, {Partial Symmetries of Weak Interactions}, Nucl. Phys. 22 (1961)
  579--588.
\newblock \href {http://dx.doi.org/10.1016/0029-5582(61)90469-2}
  {\path{doi:10.1016/0029-5582(61)90469-2}}.

\bibitem{Salam:1968rm}
A.~Salam, {Weak and Electromagnetic Interactions}, originally printed in:
  Svartholm (ed.), Elementary Particle Theory, Proceedings of the Nobel
  Symposium held 1968 at Lerum, Sweden (Stockholm 1968), p. 367-377; see also:
  \url{http://inspirehep.net/record/53083?ln=de} (1968).

\bibitem{Veltman:1977kh}
M.~Veltman, {Limit on Mass Differences in the Weinberg Model}, Nucl. Phys. B123
  (1977) 89.
\newblock \href {http://dx.doi.org/10.1016/0550-3213(77)90342-X}
  {\path{doi:10.1016/0550-3213(77)90342-X}}.

\bibitem{Sirlin:1980nh}
A.~Sirlin, {Radiative corrections in the $SU(2)_L\times U(1)$ theory: a simple
  renormalization framework}, Phys. Rev. D22 (1980) 971--981.
\newblock \href {http://dx.doi.org/10.1103/PhysRevD.22.971}
  {\path{doi:10.1103/PhysRevD.22.971}}.

\bibitem{Marciano:1980pb}
W.~J. Marciano, A.~Sirlin, Radiative corrections to neutrino induced neutral
  current phenomena in the {$SU(2)_L\times U(1)$} theory, Phys. Rev. D22 (1980)
  2695, [Erratum: Phys. Rev.D31,213(1985)].
\newblock \href {http://dx.doi.org/10.1103/PhysRevD.31.213,
  10.1103/PhysRevD.22.2695} {\path{doi:10.1103/PhysRevD.31.213,
  10.1103/PhysRevD.22.2695}}.

\bibitem{Marciano:1983wwa}
W.~J. Marciano, A.~Sirlin, Testing the {Standard Model} by precise
  determinations of {$W^{\pm}$ and $Z$} masses, Phys. Rev. D29 (1984) 945,
  [Erratum: Phys. Rev.D31,213(1985)].
\newblock \href {http://dx.doi.org/10.1103/PhysRevD.29.945,
  10.1103/PhysRevD.31.213.3} {\path{doi:10.1103/PhysRevD.29.945,
  10.1103/PhysRevD.31.213.3}}.

\bibitem{Akhundov:1985fc}
A.~Akhundov, D.~Bardin, T.~Riemann, Electroweak one loop corrections to the
  decay of the neutral vector boson, Nucl. Phys. B276 (1986) 1.
\newblock \href {http://dx.doi.org/10.1016/0550-3213(86)90014-3}
  {\path{doi:10.1016/0550-3213(86)90014-3}}.

\bibitem{Beenakker:1988pv}
W.~Beenakker, W.~Hollik, {The width of the $Z$ boson}, Z. Phys. C40 (1988) 141.
\newblock \href {http://dx.doi.org/10.1007/BF01559728}
  {\path{doi:10.1007/BF01559728}}.

\bibitem{Jegerlehner:1988ak}
F.~Jegerlehner, Precision tests of electroweak interaction parameters.~In: R.
  Manka, M. Zralek (eds.), Proc. 11$^{th}$ Int. School of Theoretical Physics,
  Testing the Standard Model, Szczyrk, Poland, Sep 18-22, 1987 (Singapore,
  World Scientific, 1988), pp. 33-108.
  \url{http://ccdb5fs.kek.jp/cgi-bin/img/allpdf?198801263}.

\bibitem{Bernabeu:1987me}
J.~Bernabeu, A.~Pich, A.~Santamaria, {$\Gamma (Z \to b{\bar b})$}: A signature
  of hard mass terms for a heavy top, Phys. Lett. B200 (1988) 569.
\newblock \href {http://dx.doi.org/10.1016/0370-2693(88)90173-6}
  {\path{doi:10.1016/0370-2693(88)90173-6}}.

\bibitem{Bardin:1986fi}
D.~Bardin, S.~Riemann, T.~Riemann, Electroweak one loop corrections to the
  decay of the charged vector boson, Z. Phys. C32 (1986) 121--125.
\newblock \href {http://dx.doi.org/10.1007/BF01441360}
  {\path{doi:10.1007/BF01441360}}.

\bibitem{Denner:1990tx}
A.~Denner, T.~Sack, {The W boson width}, Z. Phys. C46 (1990) 653--663.
\newblock \href {http://dx.doi.org/10.1007/BF01560267}
  {\path{doi:10.1007/BF01560267}}.

\bibitem{Hikasa:1992je}
K.~Hikasa, et~al., {Review of particle properties. Particle Data Group}, Phys.
  Rev. D45 (1992) S1, [Erratum: Phys. Rev. D46 (1992) 5210(1992)].
\newblock \href {http://dx.doi.org/10.1103/PhysRevD.46.5210,
  10.1103/PhysRevD.45.S1} {\path{doi:10.1103/PhysRevD.46.5210,
  10.1103/PhysRevD.45.S1}}.

\bibitem{Abe:1995hr}
F.~Abe, et~al., {Observation of top quark production in $\bar{p}p$ collisions},
  Phys. Rev. Lett. 74 (1995) 2626--2631.
\newblock \href {http://arxiv.org/abs/hep-ex/9503002}
  {\path{arXiv:hep-ex/9503002}}, \href
  {http://dx.doi.org/10.1103/PhysRevLett.74.2626}
  {\path{doi:10.1103/PhysRevLett.74.2626}}.

\bibitem{D0:1995jca}
S.~Abachi, et~al., {Observation of the top quark}, Phys. Rev. Lett. 74 (1995)
  2632--2637.
\newblock \href {http://arxiv.org/abs/hep-ex/9503003}
  {\path{arXiv:hep-ex/9503003}}, \href
  {http://dx.doi.org/10.1103/PhysRevLett.74.2632}
  {\path{doi:10.1103/PhysRevLett.74.2632}}.

\bibitem{Aad:2012tfa}
G.~Aad, et~al., {Observation of a new particle in the search for the Standard
  Model Higgs boson with the ATLAS detector at the LHC}, Phys. Lett. B716
  (2012) 1--29.
\newblock \href {http://arxiv.org/abs/1207.7214} {\path{arXiv:1207.7214}},
  \href {http://dx.doi.org/10.1016/j.physletb.2012.08.020}
  {\path{doi:10.1016/j.physletb.2012.08.020}}.

\bibitem{Djouadi:1987gn}
A.~Djouadi, C.~Verzegnassi, Virtual very heavy top effects in {LEP/SLC}
  precision measurements, Phys. Lett. B195 (1987) 265--271.
\newblock \href {http://dx.doi.org/10.1016/0370-2693(87)91206-8}
  {\path{doi:10.1016/0370-2693(87)91206-8}}.

\bibitem{Djouadi:1987di}
A.~Djouadi, ${O}(\alpha \alpha_s)$ vacuum polarization functions of the
  standard model gauge bosons, Nuovo Cim. A100 (1988) 357.
\newblock \href {http://dx.doi.org/10.1007/BF02812964}
  {\path{doi:10.1007/BF02812964}}.

\bibitem{Kniehl:1989yc}
B.~A. Kniehl, Two loop corrections to the vacuum polarizations in perturbative
  {QCD}, Nucl. Phys. B347 (1990) 86--104.
\newblock \href {http://dx.doi.org/10.1016/0550-3213(90)90552-O}
  {\path{doi:10.1016/0550-3213(90)90552-O}}.

\bibitem{Kniehl:1991gu}
B.~A. Kniehl, A.~Sirlin, {Dispersion relations for vacuum polarization
  functions in electroweak physics}, Nucl. Phys. B371 (1992) 141--148.
\newblock \href {http://dx.doi.org/10.1016/0550-3213(92)90232-Z}
  {\path{doi:10.1016/0550-3213(92)90232-Z}}.

\bibitem{Djouadi:1993ss}
A.~Djouadi, P.~Gambino, {Electroweak gauge bosons selfenergies: Complete QCD
  cor\-rec\-tions}, Phys. Rev. D49 (1994) 3499--3511, Erratum: Phys. Rev. D53
  (1996) 4111.
\newblock \href {http://arxiv.org/abs/hep-ph/9309298}
  {\path{arXiv:hep-ph/9309298}}, \href
  {http://dx.doi.org/10.1103/PhysRevD.49.3499, 10.1103/PhysRevD.53.4111}
  {\path{doi:10.1103/PhysRevD.49.3499, 10.1103/PhysRevD.53.4111}}.

\bibitem{Barbieri:1992nz}
R.~Barbieri, M.~Beccaria, P.~Ciafaloni, G.~Curci, A.~Vicere, {Radiative
  correction effects of a very heavy top}, Phys. Lett. B288 (1992) 95--98,
  [Erratum: Phys. Lett.B312,511(1993)].
\newblock \href {http://arxiv.org/abs/hep-ph/9205238}
  {\path{arXiv:hep-ph/9205238}}, \href
  {http://dx.doi.org/10.1016/0370-2693(93)90990-Y,
  10.1016/0370-2693(92)91960-H} {\path{doi:10.1016/0370-2693(93)90990-Y,
  10.1016/0370-2693(92)91960-H}}.

\bibitem{Barbieri:1992dq}
R.~Barbieri, M.~Beccaria, P.~Ciafaloni, G.~Curci, A.~Vicere, {Two loop heavy
  top effects in the Standard Model}, Nucl. Phys. B409 (1993) 105--127.
\newblock \href {http://dx.doi.org/10.1016/0550-3213(93)90448-X}
  {\path{doi:10.1016/0550-3213(93)90448-X}}.

\bibitem{Fleischer:1993ub}
J.~Fleischer, O.~V. Tarasov, F.~Jegerlehner, {Two loop heavy top corrections to
  the rho parameter: A Simple formula valid for arbitrary Higgs mass}, Phys.
  Lett. B319 (1993) 249--256.
\newblock \href {http://dx.doi.org/10.1016/0370-2693(93)90810-5}
  {\path{doi:10.1016/0370-2693(93)90810-5}}.

\bibitem{Fleischer:1994cb}
J.~Fleischer, O.~V. Tarasov, F.~Jegerlehner, {Two loop large top mass
  corrections to electroweak parameters: Analytic results valid for arbitrary
  Higgs mass}, Phys. Rev. D51 (1995) 3820--3837.
\newblock \href {http://dx.doi.org/10.1103/PhysRevD.51.3820}
  {\path{doi:10.1103/PhysRevD.51.3820}}.

\bibitem{Degrassi:1996mg}
G.~Degrassi, P.~Gambino, A.~Vicini, {Two loop heavy top effects on the
  $M_Z-M_W$ interdependence}, Phys. Lett. B383 (1996) 219--226.
\newblock \href {http://arxiv.org/abs/hep-ph/9603374}
  {\path{arXiv:hep-ph/9603374}}, \href
  {http://dx.doi.org/10.1016/0370-2693(96)00720-4}
  {\path{doi:10.1016/0370-2693(96)00720-4}}.

\bibitem{Degrassi:1996ps}
G.~Degrassi, P.~Gambino, A.~Sirlin, Precise calculation of {$M_W$},
  {$\sin^2{\hat \theta}_W(M_Z)$}, and $\sin^2 \theta_{eff}^{lept}$, Phys. Lett.
  B394 (1997) 188--194.
\newblock \href {http://arxiv.org/abs/hep-ph/9611363}
  {\path{arXiv:hep-ph/9611363}}, \href
  {http://dx.doi.org/10.1016/S0370-2693(96)01677-2}
  {\path{doi:10.1016/S0370-2693(96)01677-2}}.

\bibitem{Degrassi:1999jd}
G.~Degrassi, P.~Gambino, {Two loop heavy top corrections to the $Z^0$ boson
  partial widths}, Nucl. Phys. B567 (2000) 3--31.
\newblock \href {http://arxiv.org/abs/hep-ph/9905472}
  {\path{arXiv:hep-ph/9905472}}, \href
  {http://dx.doi.org/10.1016/S0550-3213(99)00729-4}
  {\path{doi:10.1016/S0550-3213(99)00729-4}}.

\bibitem{Freitas:2000gg}
A.~Freitas, W.~Hollik, W.~Walter, G.~Weiglein, {Complete fermionic two loop
  results for the {$M_W-M_Z$} interdependence}, Phys. Lett. B495 (2000)
  338--346, {Erratum: Phys. Lett. B570 (2003) 265}.
\newblock \href {http://arxiv.org/abs/hep-ph/0007091}
  {\path{arXiv:hep-ph/0007091}}, \href
  {http://dx.doi.org/10.1016/S0370-2693(00)01263-6,
  10.1016/j.physletb.2003.08.006} {\path{doi:10.1016/S0370-2693(00)01263-6,
  10.1016/j.physletb.2003.08.006}}.

\bibitem{Freitas:2002ja}
A.~Freitas, W.~Hollik, W.~Walter, G.~Weiglein, {Electroweak two loop
  corrections to the ${M_W - M_Z}$ mass correlation in the {Standard Model}},
  Nucl. Phys. B632 (2002) 189--218, [Erratum: Nucl. Phys. B666, 305 (2003)].
\newblock \href {http://arxiv.org/abs/hep-ph/0202131}
  {\path{arXiv:hep-ph/0202131}}, \href
  {http://dx.doi.org/10.1016/S0550-3213(02)00243-2}
  {\path{doi:10.1016/S0550-3213(02)00243-2}}.

\bibitem{Awramik:2004ge}
M.~Awramik, M.~Czakon, A.~Freitas, G.~Weiglein, {Complete two-loop electroweak
  fermionic corrections to $\sin^{2} \theta^{\rm lept}_{\rm eff}$ and indirect
  determination of the { Higgs} boson mass}, Phys. Rev. Lett. 93 (2004) 201805.
\newblock \href {http://arxiv.org/abs/hep-ph/0407317}
  {\path{arXiv:hep-ph/0407317}}, \href
  {http://dx.doi.org/10.1103/PhysRevLett.93.201805}
  {\path{doi:10.1103/PhysRevLett.93.201805}}.

\bibitem{Hollik:2005va}
W.~Hollik, U.~Meier, S.~Uccirati, {The effective electroweak mixing angle
  $\sin^2 \theta_{eff}$ with two-loop fermionic contributions}, Nucl. Phys.
  B731 (2005) 213--224.
\newblock \href {http://arxiv.org/abs/hep-ph/0507158}
  {\path{arXiv:hep-ph/0507158}}, \href
  {http://dx.doi.org/10.1016/j.nuclphysb.2005.10.015}
  {\path{doi:10.1016/j.nuclphysb.2005.10.015}}.

\bibitem{Awramik:2008gi}
M.~Awramik, M.~Czakon, A.~Freitas, B.~Kniehl, {Two-loop electroweak fermionic
  corrections to $\sin^2 \theta^{\rm b \bar b}_{\rm eff}$}, Nucl. Phys. B813
  (2009) 174--187.
\newblock \href {http://arxiv.org/abs/0811.1364} {\path{arXiv:0811.1364}},
  \href {http://dx.doi.org/10.1016/j.nuclphysb.2008.12.031}
  {\path{doi:10.1016/j.nuclphysb.2008.12.031}}.

\bibitem{Freitas:2012sy}
A.~Freitas, Y.-C. Huang, {Electroweak two-loop corrections to
  $\sin^2{\theta}_{\rm eff}^{b{\bar b}}$ and $R_{b}$ using numerical
  Mellin-Barnes integrals}, JHEP 1208 (2012) 050.
\newblock \href {http://arxiv.org/abs/1205.0299} {\path{arXiv:1205.0299}},
  \href {http://dx.doi.org/10.1007/JHEP08(2012)050}
  {\path{doi:10.1007/JHEP08(2012)050}}.

\bibitem{Freitas:2013dpa}
A.~Freitas, {Two-loop fermionic electroweak corrections to the Z-boson width
  and production rate}, Phys. Lett. B730 (2014) 50--52.
\newblock \href {http://arxiv.org/abs/1310.2256} {\path{arXiv:1310.2256}},
  \href {http://dx.doi.org/10.1016/j.physletb.2014.01.017}
  {\path{doi:10.1016/j.physletb.2014.01.017}}.

\bibitem{Freitas:2014hra}
A.~Freitas, {Higher-order electroweak corrections to the partial widths and
  branching ratios of the Z boson}, JHEP 1404 (2014) 070.
\newblock \href {http://arxiv.org/abs/1401.2447} {\path{arXiv:1401.2447}},
  \href {http://dx.doi.org/10.1007/JHEP04(2014)070}
  {\path{doi:10.1007/JHEP04(2014)070}}.

\bibitem{Avdeev:1994db}
L.~Avdeev, J.~Fleischer, S.~Mikhailov, O.~Tarasov, {${O}( \alpha \alpha_s^2 )$
  correction to the electroweak $\rho$ parameter}, Phys. Lett. B336 (1994)
  560--566, {Erratum: Phys. Lett. B349 (1995) 597}.
\newblock \href {http://arxiv.org/abs/hep-ph/9406363}
  {\path{arXiv:hep-ph/9406363}}, \href
  {http://dx.doi.org/10.1016/0370-2693(94)90573-8}
  {\path{doi:10.1016/0370-2693(94)90573-8}}.

\bibitem{Chetyrkin:1995ix}
K.~Chetyrkin, J.~H. K{\"u}hn, M.~Steinhauser, {Corrections of order ${O}(G_F
  M_t^2 \alpha_s^2)$ to the $\rho$ parameter}, Phys. Lett. B351 (1995)
  331--338.
\newblock \href {http://arxiv.org/abs/hep-ph/9502291}
  {\path{arXiv:hep-ph/9502291}}, \href
  {http://dx.doi.org/10.1016/0370-2693(95)00380-4}
  {\path{doi:10.1016/0370-2693(95)00380-4}}.

\bibitem{vanderBij:2000cg}
J.~J. van~der Bij, K.~G. Chetyrkin, M.~Faisst, G.~Jikia, T.~Seidensticker,
  {Three loop leading top mass contributions to the $\rho$ parameter}, Phys.
  Lett. B498 (2001) 156--162.
\newblock \href {http://arxiv.org/abs/hep-ph/0011373}
  {\path{arXiv:hep-ph/0011373}}, \href
  {http://dx.doi.org/10.1016/S0370-2693(01)00002-8}
  {\path{doi:10.1016/S0370-2693(01)00002-8}}.

\bibitem{Faisst:2003px}
M.~Faisst, J.~H. K{\"u}hn, T.~Seidensticker, O.~Veretin, {Three loop top quark
  contributions to the $\rho$ parameter}, Nucl. Phys. B665 (2003) 649--662.
\newblock \href {http://arxiv.org/abs/hep-ph/0302275}
  {\path{arXiv:hep-ph/0302275}}, \href
  {http://dx.doi.org/10.1016/S0550-3213(03)00450-4}
  {\path{doi:10.1016/S0550-3213(03)00450-4}}.

\bibitem{Schroder:2005db}
Y.~Schr{\"o}der, M.~Steinhauser, {Four-loop singlet contribution to the $\rho$
  parameter}, Phys. Lett. B622 (2005) 124--130.
\newblock \href {http://arxiv.org/abs/hep-ph/0504055}
  {\path{arXiv:hep-ph/0504055}}, \href
  {http://dx.doi.org/10.1016/j.physletb.2005.06.085}
  {\path{doi:10.1016/j.physletb.2005.06.085}}.

\bibitem{Chetyrkin:2006bj}
K.~G. Chetyrkin, M.~Faisst, J.~H. K{\"u}hn, P.~Maierhofer, C.~Sturm, Four-loop
  {QCD} corrections to the $\rho$ parameter, Phys. Rev. Lett. 97 (2006) 102003.
\newblock \href {http://arxiv.org/abs/hep-ph/0605201}
  {\path{arXiv:hep-ph/0605201}}, \href
  {http://dx.doi.org/10.1103/PhysRevLett.97.102003}
  {\path{doi:10.1103/PhysRevLett.97.102003}}.

\bibitem{Boughezal:2006xk}
R.~Boughezal, M.~Czakon, {Single scale tadpoles and $O(G_F m_t^2 \alpha_s^3)$
  corrections to the $\rho$ parameter}, Nucl. Phys. B755 (2006) 221--238.
\newblock \href {http://arxiv.org/abs/hep-ph/0606232}
  {\path{arXiv:hep-ph/0606232}}, \href
  {http://dx.doi.org/10.1016/j.nuclphysb.2006.08.007}
  {\path{doi:10.1016/j.nuclphysb.2006.08.007}}.

\bibitem{Awramik:2002wn}
M.~Awramik, M.~Czakon, {Complete two loop bosonic contributions to the muon
  lifetime in the Standard Model}, Phys. Rev. Lett. 89 (2002) 241801.
\newblock \href {http://arxiv.org/abs/hep-ph/0208113}
  {\path{arXiv:hep-ph/0208113}}, \href
  {http://dx.doi.org/10.1103/PhysRevLett.89.241801}
  {\path{doi:10.1103/PhysRevLett.89.241801}}.

\bibitem{Awramik:2003ee}
M.~Awramik, M.~Czakon, {Complete two loop electroweak contributions to the muon
  lifetime in the Standard Model}, Phys. Lett. B568 (2003) 48--54.
\newblock \href {http://arxiv.org/abs/hep-ph/0305248}
  {\path{arXiv:hep-ph/0305248}}, \href
  {http://dx.doi.org/10.1016/j.physletb.2003.06.007}
  {\path{doi:10.1016/j.physletb.2003.06.007}}.

\bibitem{Onishchenko:2002ve}
A.~Onishchenko, O.~Veretin, {Two loop bosonic electroweak corrections to the
  muon lifetime and ${M_Z - M_W}$ interdependence}, Phys. Lett. B551 (2003)
  111--114.
\newblock \href {http://arxiv.org/abs/hep-ph/0209010}
  {\path{arXiv:hep-ph/0209010}}, \href
  {http://dx.doi.org/10.1016/S0370-2693(02)03004-6}
  {\path{doi:10.1016/S0370-2693(02)03004-6}}.

\bibitem{Awramik:2003rn}
M.~Awramik, M.~Czakon, A.~Freitas, G.~Weiglein, {Precise prediction for the {W}
  boson mass in the Standard Model}, Phys. Rev. D69 (2004) 053006.
\newblock \href {http://arxiv.org/abs/hep-ph/0311148}
  {\path{arXiv:hep-ph/0311148}}, \href
  {http://dx.doi.org/10.1103/PhysRevD.69.053006}
  {\path{doi:10.1103/PhysRevD.69.053006}}.

\bibitem{Awramik:2006ar}
M.~Awramik, M.~Czakon, A.~Freitas, {Bosonic corrections to the effective weak
  mixing angle at $O(\alpha^2)$}, Phys. Lett. B642 (2006) 563--566.
\newblock \href {http://arxiv.org/abs/hep-ph/0605339}
  {\path{arXiv:hep-ph/0605339}}, \href
  {http://dx.doi.org/10.1016/j.physletb.2006.07.035}
  {\path{doi:10.1016/j.physletb.2006.07.035}}.

\bibitem{Hollik:2006ma}
W.~Hollik, U.~Meier, S.~Uccirati, {The effective electroweak mixing angle
  $\sin^2 \theta_{eff}$ with two-loop bosonic contributions}, Nucl. Phys. B765
  (2007) 154--165.
\newblock \href {http://arxiv.org/abs/hep-ph/0610312}
  {\path{arXiv:hep-ph/0610312}}, \href
  {http://dx.doi.org/10.1016/j.nuclphysb.2006.12.001}
  {\path{doi:10.1016/j.nuclphysb.2006.12.001}}.

\bibitem{Awramik:2006uz}
M.~Awramik, M.~Czakon, A.~Freitas, {Electroweak two-loop corrections to the
  effective weak mixing angle}, JHEP 11 (2006) 048.
\newblock \href {http://arxiv.org/abs/hep-ph/0608099}
  {\path{arXiv:hep-ph/0608099}}, \href
  {http://dx.doi.org/10.1088/1126-6708/2006/11/048}
  {\path{doi:10.1088/1126-6708/2006/11/048}}.

\bibitem{Dubovyk:2016aqv}
I.~Dubovyk, A.~Freitas, J.~Gluza, T.~Riemann, J.~Usovitsch, {The two-loop
  electroweak bosonic corrections to $\sin^2 \theta_{\mathrm{eff}}^b$}, Phys.
  Lett. B762 (2016) 184--189.
\newblock \href {http://arxiv.org/abs/1607.08375} {\path{arXiv:1607.08375}},
  \href {http://dx.doi.org/10.1016/j.physletb.2016.09.012}
  {\path{doi:10.1016/j.physletb.2016.09.012}}.

\bibitem{Bardin:1988xt}
D.~Bardin, A.~Leike, T.~Riemann, M.~Sachwitz, Energy dependent width effects in
  $e^+ e^-$ annihilation near the {Z} boson pole, Phys. Lett. B206 (1988)
  539--542.
\newblock \href {http://dx.doi.org/10.1016/0370-2693(88)91625-5}
  {\path{doi:10.1016/0370-2693(88)91625-5}}.

\bibitem{Chetyrkin:1994js}
K.~Chetyrkin, J.~H. K{\"u}hn, A.~Kwiatkowski, {QCD corrections to the $e^{+}
  e^{-}$ cross-section and the $Z$ boson decay rate.~}In: Reports of the
  working group on precision calculations for the Z resonance, p. 175-263
  (Geneva 1994), CERN 95-03 ( 31 March 1995). See also: QCD corrections to the
  $e^+e^-$ cross-section and the {Z} boson decay rate: Concepts and results,
  March 1996, 105 pages, LBL-36678-REV and {\bf Phys. Rept.} 277 (1996)
  189-281.
\newblock \href {http://arxiv.org/abs/hep-ph/9503396}
  {\path{arXiv:hep-ph/9503396}}.

\bibitem{Baikov:2008jh}
P.~Baikov, K.~Chetyrkin, J.~K{\"u}hn, {Order $\alpha^4 (s)$ {QCD} Corrections
  to {Z} and $\tau$ Decays}, Phys. Rev. Lett. 101 (2008) 012002.
\newblock \href {http://arxiv.org/abs/0801.1821} {\path{arXiv:0801.1821}},
  \href {http://dx.doi.org/10.1103/PhysRevLett.101.012002}
  {\path{doi:10.1103/PhysRevLett.101.012002}}.

\bibitem{Kataev:1992dg}
A.~L. Kataev, {Higher order $O(\alpha^2)$ and $O(\alpha \alpha_s)$ corrections
  to $\sigma_{tot}(e^+ e^- \to {\mathrm{hadrons}})$ and Z boson decay rate},
  Phys. Lett. B287 (1992) 209--212.
\newblock \href {http://dx.doi.org/10.1016/0370-2693(92)91901-K}
  {\path{doi:10.1016/0370-2693(92)91901-K}}.

\bibitem{Steinhauser:1998rq-new}
M.~Steinhauser, {Leptonic contribution to the effective electromagnetic
  coupling constant up to three loops}, Phys. Lett. B429 (1998) 158--161.
\newblock \href {http://arxiv.org/abs/hep-ph/9803313}
  {\path{arXiv:hep-ph/9803313}}, \href
  {http://dx.doi.org/10.1016/S0370-2693(98)00503-6}
  {\path{doi:10.1016/S0370-2693(98)00503-6}}.

\bibitem{Davier:2017zfy}
M.~Davier, A.~Hoecker, B.~Malaescu, Z.~Zhang, {Reevaluation of the hadronic
  vacuum polarisation contributions to the Standard Model predictions of the
  muon $g$-2 and ${\alpha (m_Z^2)}$ using newest hadronic cross-section data},
  Eur. Phys. J. C77~(12) (2017) 827.
\newblock \href {http://arxiv.org/abs/1706.09436} {\path{arXiv:1706.09436}},
  \href {http://dx.doi.org/10.1140/epjc/s10052-017-5161-6}
  {\path{doi:10.1140/epjc/s10052-017-5161-6}}.

\bibitem{Jegerlehner:2017zsb}
F.~Jegerlehner, {Variations on Photon Vacuum Polarization.~}\href
  {http://arxiv.org/abs/1711.06089} {\path{arXiv:1711.06089}}.

\bibitem{Keshavarzi:2018mgv}
A.~Keshavarzi, D.~Nomura, T.~Teubner, {The muon $g$-2 and $\alpha(M_Z^2)$: a
  new data-based analysis.~}\href {http://arxiv.org/abs/1802.02995}
  {\path{arXiv:1802.02995}}.

\bibitem{Dubovyk:2016ocz}
I.~Dubovyk, J.~Gluza, T.~Riemann, J.~Usovitsch, {Numerical integration of
  massive two-loop Mellin-Barnes integrals in Minkowskian regions}, PoS LL2016
  (2016) 034, \url{https://pos.sissa.it/260/034/pdf}.
\newblock \href {http://arxiv.org/abs/1607.07538} {\path{arXiv:1607.07538}}.

\bibitem{Dubovyk:2016zok}
I.~Dubovyk, A.~Freitas, J.~Gluza, T.~Riemann, J.~Usovitsch, {30 years, some 700
  integrals, and 1 dessert, or: Electroweak two-loop corrections to the Z$\bar
  b$b vertex}, PoS LL2016 (2016) 075, \url{https://pos.sissa.it/260/075/pdf}.
\newblock \href {http://arxiv.org/abs/1610.07059} {\path{arXiv:1610.07059}}.

\bibitem{Dubovyk:2017cqw}
I.~Dubovyk, J.~Gluza, T.~Jelinski, T.~Riemann, J.~Usovitsch, {New prospects for
  the numerical calculation of Mellin-Barnes integrals in Minkowskian
  kinematics}, Acta Phys. Polon. B48 (2017) 995.
\newblock \href {http://arxiv.org/abs/1704.02288} {\path{arXiv:1704.02288}},
  \href {http://dx.doi.org/10.5506/APhysPolB.48.995}
  {\path{doi:10.5506/APhysPolB.48.995}}.

\bibitem{Hahn:2000kx}
T.~Hahn, {Generating Feynman diagrams and amplitudes with FeynArts 3}, Comput.
  Phys. Commun. 140 (2001) 418--431.
\newblock \href {http://arxiv.org/abs/hep-ph/0012260}
  {\path{arXiv:hep-ph/0012260}}, \href
  {http://dx.doi.org/10.1016/S0010-4655(01)00290-9}
  {\path{doi:10.1016/S0010-4655(01)00290-9}}.

\bibitem{Bauberger:1994by}
S.~Bauberger, F.~A. Berends, M.~B{\"o}hm, M.~Buza, {Analytical and numerical
  methods for massive two loop selfenergy diagrams}, Nucl. Phys. B434 (1995)
  383--407.
\newblock \href {http://arxiv.org/abs/hep-ph/9409388}
  {\path{arXiv:hep-ph/9409388}}, \href
  {http://dx.doi.org/10.1016/0550-3213(94)00475-T}
  {\path{doi:10.1016/0550-3213(94)00475-T}}.

\bibitem{Bauberger:1994hx}
S.~Bauberger, M.~B{\"o}hm, {Simple one-dimensional integral representations for
  two loop selfenergies: The master diagram}, Nucl. Phys. B445 (1995) 25--48.
\newblock \href {http://arxiv.org/abs/hep-ph/9501201}
  {\path{arXiv:hep-ph/9501201}}, \href
  {http://dx.doi.org/10.1016/0550-3213(95)00199-3}
  {\path{doi:10.1016/0550-3213(95)00199-3}}.

\bibitem{Hepp:1966eg}
K.~Hepp, {Proof of the Bogolyubov-Parasiuk theorem on renormalization}, Commun.
  Math. Phys. 2 (1966) 301--326.
\newblock \href {http://dx.doi.org/10.1007/BF01773358}
  {\path{doi:10.1007/BF01773358}}.

\bibitem{Binoth:2000ps}
T.~Binoth, G.~Heinrich, {An automatized algorithm to compute infrared divergent
  multiloop integrals}, Nucl. Phys. B585 (2000) 741--759.
\newblock \href {http://arxiv.org/abs/hep-ph/0004013}
  {\path{arXiv:hep-ph/0004013}}, \href
  {http://dx.doi.org/10.1016/S0550-3213(00)00429-6}
  {\path{doi:10.1016/S0550-3213(00)00429-6}}.

\bibitem{Borowka:2015mxa}
S.~Borowka, G.~Heinrich, S.~P. Jones, M.~Kerner, J.~Schlenk, T.~Zirke, {SecDec
  3.0: numerical evaluation of multi-scale integrals beyond one loop}, Comput.
  Phys. Commun. 196 (2015) 470--491.
\newblock \href {http://arxiv.org/abs/1502.06595} {\path{arXiv:1502.06595}},
  \href {http://dx.doi.org/10.1016/j.cpc.2015.05.022}
  {\path{doi:10.1016/j.cpc.2015.05.022}}.

\bibitem{Smirnov:2013eza}
A.~V. Smirnov, {FIESTA 3: cluster-parallelizable multiloop numerical
  calculations in physical regions}, Comput. Phys. Commun. 185 (2014)
  2090--2100.
\newblock \href {http://arxiv.org/abs/1312.3186} {\path{arXiv:1312.3186}},
  \href {http://dx.doi.org/10.1016/j.cpc.2014.03.015}
  {\path{doi:10.1016/j.cpc.2014.03.015}}.

\bibitem{Usyukina:1975yg}
N.~I. Usyukina, On a representation for three point function, Teor. Mat. Fiz.
  22 (1975) 300--306.
\newblock \href {http://dx.doi.org/10.1007/BF01037795}
  {\path{doi:10.1007/BF01037795}}.

\bibitem{Smirnov:1999gc}
V.~A. Smirnov, {Analytical result for dimensionally regularized massless on
  shell double box}, Phys. Lett. B460 (1999) 397--404.
\newblock \href {http://arxiv.org/abs/hep-ph/9905323}
  {\path{arXiv:hep-ph/9905323}}, \href
  {http://dx.doi.org/10.1016/S0370-2693(99)00777-7}
  {\path{doi:10.1016/S0370-2693(99)00777-7}}.

\bibitem{Tausk:1999vh}
J.~Tausk, {Nonplanar massless two loop Feynman diagrams with four on-shell
  legs}, Phys. Lett. B469 (1999) 225--234.
\newblock \href {http://arxiv.org/abs/hep-ph/9909506}
  {\path{arXiv:hep-ph/9909506}}, \href
  {http://dx.doi.org/10.1016/S0370-2693(99)01277-0}
  {\path{doi:10.1016/S0370-2693(99)01277-0}}.

\bibitem{mbtools}
M. Czakon (MB, MBasymptotics), D. Kosower (barnesroutines), A. Smirnov, V.
  Smirnov (MBresolve), K. Bielas, I. Dubovyk, J. Gluza, K. Kajda, T. Riemann
  (AMBRE, PlanarityTest), MBtools webpage, \url{https://mbtools.hepforge.org/}.

\bibitem{Czakon:2005rk}
M.~Czakon, {Automatized analytic continuation of Mellin-Barnes integrals},
  Comput. Phys. Commun. 175 (2006) 559--571.
\newblock \href {http://arxiv.org/abs/hep-ph/0511200}
  {\path{arXiv:hep-ph/0511200}}, \href
  {http://dx.doi.org/10.1016/j.cpc.2006.07.002}
  {\path{doi:10.1016/j.cpc.2006.07.002}}.

\bibitem{Smirnov:2009up}
A.~Smirnov, V.~Smirnov, {On the Resolution of Singularities of Multiple
  Mellin-Barnes Integrals}, Eur. Phys. J. C62 (2009) 445--449.
\newblock \href {http://arxiv.org/abs/0901.0386} {\path{arXiv:0901.0386}},
  \href {http://dx.doi.org/10.1140/epjc/s10052-009-1039-6}
  {\path{doi:10.1140/epjc/s10052-009-1039-6}}.

\bibitem{Gluza:2007rt}
J.~Gluza, K.~Kajda, T.~Riemann, {AMBRE - a Mathematica package for the
  construction of Mellin-Barnes representations for Feynman integrals}, Comput.
  Phys. Commun. 177 (2007) 879--893.
\newblock \href {http://arxiv.org/abs/0704.2423} {\path{arXiv:0704.2423}},
  \href {http://dx.doi.org/10.1016/j.cpc.2007.07.001}
  {\path{doi:10.1016/j.cpc.2007.07.001}}.

\bibitem{Bielas:2013rja}
K.~Bielas, I.~Dubovyk, J.~Gluza, T.~Riemann, {Some Remarks on Non-planar
  Feynman Diagrams}, Acta Phys. Polon. B44 (2013) 2249--2255.
\newblock \href {http://arxiv.org/abs/1312.5603} {\path{arXiv:1312.5603}},
  \href {http://dx.doi.org/10.5506/APhysPolB.44.2249}
  {\path{doi:10.5506/APhysPolB.44.2249}}.

\bibitem{Gluza:2010rn}
J.~Gluza, K.~Kajda, T.~Riemann, V.~Yundin, {Numerical Evaluation of Tensor
  Feynman Integrals in Euclidean Kinematics}, Eur. Phys. J. C71 (2011) 1516.
\newblock \href {http://arxiv.org/abs/1010.1667} {\path{arXiv:1010.1667}},
  \href {http://dx.doi.org/10.1140/epjc/s10052-010-1516-y}
  {\path{doi:10.1140/epjc/s10052-010-1516-y}}.

\bibitem{Dubovyk:2015yba}
I.~Dubovyk, J.~Gluza, T.~Riemann, {Non-planar Feynman diagrams and
  Mellin-Barnes representations with $\small{AMBRE}$ $\small{3.0}$}, J. Phys.
  Conf. Ser. 608 (2015) 012070. DESY 14--174 (13 April 2016).
\newblock \href {http://dx.doi.org/10.1088/1742-6596/608/1/012070}
  {\path{doi:10.1088/1742-6596/608/1/012070}}.

\bibitem{Ochman:2015fho}
M.~Ochman, T.~Riemann, {MBsums - a Mathematica package for the representation
  of Mellin-Barnes integrals by multiple sums}, Acta Phys. Polon. B46 (2015)
  2117.
\newblock \href {http://arxiv.org/abs/1511.01323} {\path{arXiv:1511.01323}},
  \href {http://dx.doi.org/10.5506/APhysPolB.46.2117}
  {\path{doi:10.5506/APhysPolB.46.2117}}.

\bibitem{Katowice-CAS:2007}
Silesian University at Katowice, webpage
  \url{http://prac.us.edu.pl/~gluza/ambre/}.

\bibitem{Usovitsch:201606}
I.~Dubovyk, T.~Riemann, J.~Usovitsch, {Numerical calculation of multiple
  MB-integral representations for Feynman integrals, webpage
  \url{http://prac.us.edu.pl/~gluza/ambre/}. J. Usovitsch, Mathematica/C++
  package MBnumerics, to appear.}

\bibitem{Hahn:2004fe}
T.~Hahn, {CUBA: A library for multidimensional numerical integration}, Comput.
  Phys. Commun. 168 (2005) 78--95.
\newblock \href {http://arxiv.org/abs/hep-ph/0404043}
  {\path{arXiv:hep-ph/0404043}}, \href
  {http://dx.doi.org/10.1016/j.cpc.2005.01.010}
  {\path{doi:10.1016/j.cpc.2005.01.010}}.

\bibitem{Hahn:2014fua}
T.~Hahn, {Concurrent CUBA}, J. Phys. Conf. Ser. 608 (2015) 012066.
\newblock \href {http://arxiv.org/abs/1408.6373} {\path{arXiv:1408.6373}},
  \href {http://dx.doi.org/10.1088/1742-6596/608/1/012066}
  {\path{doi:10.1088/1742-6596/608/1/012066}}.

\bibitem{Czarnecki:1996ei}
A.~Czarnecki, J.~H. K{\"u}hn, {Nonfactorizable QCD and electroweak corrections
  to the hadronic Z boson decay rate}, Phys. Rev. Lett. 77 (1996) 3955--3958.
\newblock \href {http://arxiv.org/abs/hep-ph/9608366}
  {\path{arXiv:hep-ph/9608366}}, \href
  {http://dx.doi.org/10.1103/PhysRevLett.77.3955}
  {\path{doi:10.1103/PhysRevLett.77.3955}}.

\bibitem{Harlander:1997zb}
R.~Harlander, T.~Seidensticker, M.~Steinhauser, {Complete corrections of order
  $O(\alpha \alpha_s)$ to the decay of the Z boson into bottom quarks}, Phys.
  Lett. B426 (1998) 125--132.
\newblock \href {http://arxiv.org/abs/hep-ph/9712228}
  {\path{arXiv:hep-ph/9712228}}, \href
  {http://dx.doi.org/10.1016/S0370-2693(98)00220-2}
  {\path{doi:10.1016/S0370-2693(98)00220-2}}.

\bibitem{Fleischer:1992fq}
J.~Fleischer, O.~V. Tarasov, F.~Jegerlehner, P.~Raczka, {Two loop $O (\alpha_s
  G_{\mu} m_{t}^2)$ corrections to the partial decay width of the $Z^0$ into
  $b{\bar b}$ final states in the large top mass limit}, Phys. Lett. B293
  (1992) 437--444.
\newblock \href {http://dx.doi.org/10.1016/0370-2693(92)90909-N}
  {\path{doi:10.1016/0370-2693(92)90909-N}}.

\bibitem{Buchalla:1992zm}
G.~Buchalla, A.~J. Buras, {QCD corrections to the $\bar s d Z$ vertex for
  arbitrary top quark mass}, Nucl. Phys. B398 (1993) 285--300.
\newblock \href {http://dx.doi.org/10.1016/0550-3213(93)90110-B}
  {\path{doi:10.1016/0550-3213(93)90110-B}}.

\bibitem{Degrassi:1993ij}
G.~Degrassi, {Current algebra approach to heavy top effects in $Z \to b + {\bar
  b}$}, Nucl. Phys. B407 (1993) 271--289.
\newblock \href {http://arxiv.org/abs/hep-ph/9302288}
  {\path{arXiv:hep-ph/9302288}}, \href
  {http://dx.doi.org/10.1016/0550-3213(93)90058-W}
  {\path{doi:10.1016/0550-3213(93)90058-W}}.

\bibitem{Chetyrkin:1993jp}
K.~Chetyrkin, A.~Kwiatkowski, M.~Steinhauser, {Leading top mass corrections of
  order ${O (\alpha \alpha_s m_t^2 / M_W^2)}$ to partial decay rate {$\Gamma (Z
  \to b {\bar b})$}}, Mod. Phys. Lett. A8 (1993) 2785--2792,
  \href{http://dx.doi.org/10.1142/S0217732393003172}{doi:10.1142/S0217732393003172}.
\newblock \href {http://dx.doi.org/10.1142/S0217732393003172}
  {\path{doi:10.1142/S0217732393003172}}.

\bibitem{Patrignani:2016xqp}
C.~Patrignani, et~al., {Review of Particle Physics}, Chin. Phys. C40~(10)
  (2016) 100001.
\newblock \href {http://dx.doi.org/10.1088/1674-1137/40/10/100001}
  {\path{doi:10.1088/1674-1137/40/10/100001}}.

\bibitem{Freitas:2016sty}
A.~Freitas, {Numerical multi-loop integrals and applications}, Prog. Part.
  Nucl. Phys. 90 (2016) 201--240.
\newblock \href {http://arxiv.org/abs/1604.00406} {\path{arXiv:1604.00406}},
  \href {http://dx.doi.org/10.1016/j.ppnp.2016.06.004}
  {\path{doi:10.1016/j.ppnp.2016.06.004}}.

\bibitem{Bardin:1999yd}
D.~Bardin, P.~Christova, M.~Jack, L.~Kalinovskaya, A.~Olchevski, S.~Riemann,
  T.~Riemann, {ZFITTER v.6.21: A semianalytical program for fermion pair
  production in $e^+ e^-$ annihilation}, Comput. Phys. Commun. 133 (2001)
  229--395.
\newblock \href {http://arxiv.org/abs/hep-ph/9908433}
  {\path{arXiv:hep-ph/9908433}}, \href
  {http://dx.doi.org/10.1016/S0010-4655(00)00152-1}
  {\path{doi:10.1016/S0010-4655(00)00152-1}}.

\bibitem{Arbuzov:2005ma}
A.~Arbuzov, M.~Awramik, M.~Czakon, A.~Freitas, M.~Gr{\"u}newald, K.~M{\"o}nig,
  S.~Riemann, T.~Riemann, {ZFITTER: A semi-analytical program for fermion pair
  production in $e^+ e^-$ annihilation, from version 6.21 to version 6.42},
  Comput. Phys. Commun. 174 (2006) 728--758.
\newblock \href {http://arxiv.org/abs/hep-ph/0507146}
  {\path{arXiv:hep-ph/0507146}}, \href
  {http://dx.doi.org/10.1016/j.cpc.2005.12.009}
  {\path{doi:10.1016/j.cpc.2005.12.009}}.

\bibitem{Akhundov:2013ons}
A.~Akhundov, A.~Arbuzov, S.~Riemann, T.~Riemann, {The ZFITTER project}, Phys.
  Part. Nucl. 45~(3) (2014) 529--549.
\newblock \href {http://arxiv.org/abs/1302.1395} {\path{arXiv:1302.1395}},
  \href {http://dx.doi.org/10.1134/S1063779614030022}
  {\path{doi:10.1134/S1063779614030022}}.

\bibitem{Bardin:1999gt}
D.~Bardin, M.~Gr{\"u}newald, G.~Passarino, {Precision calculation project
  report} (1999).
\newblock \href {http://arxiv.org/abs/hep-ph/9902452}
  {\path{arXiv:hep-ph/9902452}}.

\bibitem{Leike:1991pq}
A.~Leike, T.~Riemann, J.~Rose, {S matrix approach to the Z line shape}, Phys.
  Lett. B273 (1991) 513--518.
\newblock \href {http://arxiv.org/abs/hep-ph/9508390}
  {\path{arXiv:hep-ph/9508390}}, \href
  {http://dx.doi.org/10.1016/0370-2693(91)90307-C}
  {\path{doi:10.1016/0370-2693(91)90307-C}}.

\bibitem{Riemann:1992gv}
T.~Riemann, {Cross-section asymmetries around the Z peak}, Phys. Lett. B293
  (1992) 451--456.
\newblock \href {http://arxiv.org/abs/hep-ph/9506382}
  {\path{arXiv:hep-ph/9506382}}, \href
  {http://dx.doi.org/10.1016/0370-2693(92)90911-M}
  {\path{doi:10.1016/0370-2693(92)90911-M}}.

\bibitem{Riemann:2015wpn}
T.~Riemann, {S-matrix Approach to the $Z$ Resonance}, Acta Phys. Polon.
  B46~(11) (2015) 2235.
\newblock \href {http://arxiv.org/abs/1610.04501} {\path{arXiv:1610.04501}},
  \href {http://dx.doi.org/10.5506/APhysPolB.46.2235}
  {\path{doi:10.5506/APhysPolB.46.2235}}.

\bibitem{QED-3loops}
T.~Riemann et al., Unfolding real observables at the $Z$ peak, to appear as
  contribution to the report on \cite{mini},
  \url{https://indico.cern.ch/event/669224/contributions/2805424/attachments/1581663/2499614/2018-cern-ceex.pdf}
  and \url{
  
https://indico.cern.ch/event/669224/contributions/2805418/attachments/1581648/2504620/riemann-FCCeeminiCERN-short.pdf}.

\end{thebibliography}

\end{document}